\documentclass[ 10pt, journal = mamobx, manuscript = article ]{achemso}

\usepackage[T1]{fontenc} 
\usepackage[utf8]{inputenc}
\usepackage[T1]{fontenc}
\usepackage{graphicx,amsmath,leftidx,amsfonts, mathtools}
\usepackage{subfig,array,booktabs,diagbox,chemfig}
\usepackage[version=4]{mhchem}
\usepackage{color,soul}
\setstcolor{red}
\usepackage{caption}
\author{Luca Banetta}
\affiliation{Department of Chemical
Engineering and Biotechnology, University of Cambridge, Cambridge, CB3 0AS, U.K.}
\email{lb725@cam.ac.uk}

\author{Giuseppe Storti}
\affiliation{Dipartimento di Chimica, Materiali ed Ingegneria Chimica "Giulio Natta", Politecnico di Milano, Italy}

\author{George Hoggard}
\affiliation{Synthomer (UK) Limited, South Marsh Road, Grimsby DN41 8DB, U.K.}

\author{Gareth Simpson}
\affiliation{Synthomer (UK) Limited, Central Road, CM20 2BH Harlow, U.K.}

\author{Alessio Zaccone}
\affiliation{Dipartimento di Fisica "A. Pontremoli", University of Milan, via Celoria 16, 20133 Milan, Italy}
\email{alessio.zaccone@unimi.it}

\title{Predictive model of polymer reaction kinetics and coagulation behavior in seeded emulsion co- and ter-polymerizations}

\abbreviations{IR,NMR,UV}

\begin{document}
	






\begin{abstract}
A mathematical model to describe the emulsion polymerization kinetics of co- and ter-polymerizations is developed. The model is based on the classical Smith-Ewart (SE) equations, within the pseudo-homopolymerization approach, with state-of-the-art models for radical entry and desorption. For co- and ter-polymerizations there are unknown parameters in the model which  are related to monomer-specific gel-effect coefficients, that are needed to compute the bimolecular termination reaction rates. The unknown parameters are determined through extensive calibration of the model on literature data for homo- and co-polymerizations
of \textit{n}-butyl acrylate (n-BA) and methyl methacrylate (MMA). The so-obtained predictive model is then applied to the modelling of the ter-polymerization of n-BA and MMA with 2-hydroxyethyl methacrylate (2-HEMA) with sodium persulphate (SPR) as initiator: predictions for the time-evolution of particle size and conversion are in excellent agreement with experimental measurements using Dynamic Light Scattering (DLS) and Gas Chromatography (GC), upon tuning the gel-effect coefficient related to 2-HEMA. The developed model is used to quantify the surfactant surface coverage of the particles as well as the total concentration of counterions in the system throughout the entire polymerization process. This key information provides a way to rationalize and control the coagulation behavior during the whole polymerization process.

\end{abstract}
\section{Introduction}
Emulsion polymerization is one the most popular processes for the synthesis and large-scale production of a great variety of polymers in colloidal form (latex), with a broad application range, including adhesives, paints, medical materials and additives for paper \cite{Thickett}.
This polymerization technique has a number of advantages when compared to other methods such as suspension and bulk polymerizations. The use of water as liquid medium instead of organic solvents is more gentle towards the environment and eases the removal of the heat produced during the reaction. Moreover, it guarantees the possibility to obtain waterborne dispersions with a solid content over 50\%, a feature highly desirable for many products, given its lower costs for transport and the faster medium evaporation.\\
One of the major issues when it comes to the production of latexes with such high values of solid content is the possibility of considerable reactor fouling due to the appearance of coagulum (coagulation of the polymer colloid).
It is the result of uncontrolled aggregation which leads to the formation of both microscopic and macroscopic agglomerates due to which the final product cannot achieve the required design features. This may lead to considerable economic losses which may have a substantial impact on the profitability of industrial emulsion polymerization processes.\\
There are two major causes which trigger the coagulation kinetics \cite{Vanderhoff}: (i) Solid phase subject to strong shear rates inside the reactor; (ii) Loss of colloidal stability. 
There are of course other phenomena which can cause formation of coagulum: one of them is the secondary nucleation, but this mechanism is non negligible for number concentration of particles $[N_P] \approx 10^{14} \text{L}^{-1}$ \cite{Gilbert_Review} or lower, meanwhile in this work the test cases used for analysis of coagulum foresee values of $[N_P]$ at least two orders of magnitude higher, so this mechanism will not be considered at this stage of the project.\\
A considerable amount of work has been focused on mechanism (i), which causes the so called \textit{mechanical coagulum}. Mat{\v{e}}j{\'\i}{\v{c}}ek and co-workers \cite{Matejicek} have observed a dual influence of the reactor agitation on the appearance of coagulum during the ter-polymerization of styrene/butyl acrylate/acrylic acid: an initial increase of power provided to the impeller decreases the fouling thanks to a better mixing but, above a certain power threshold, the coagulum increases because of the increasingly important contribution of the shear-induced aggregation process. Lowry et al. \cite{Lowry} have built a semi-empirical model which predicts an increase of fouling with increasing power provided to the impeller, because more frequent collisions become capable of overcoming the energy barrier between the particles.\\
The second mechanism (ii), the loss of colloidal stability, has been widely studied as well. Zubitur et al. \cite{Zubitur} observed a considerable amount of coagulum during the polymerization of a styrene-butyl acrylate co-polymer because of poor mixing conditions: they observed a reduction of coagulum with the increase of rotational speed of the impeller because of a reduction of the size of stagnant zones due to better mixing, especially next to  the shaft and the liquid-air interface, the most common loci of coagulum formation.\\
Even if a complete and quantitative understanding of such phenomena is quite difficult, it can be said that the impelling power during a polymerization should not be too low, to ensure sufficient homogeneous mixing conditions inside the reactor, but at the same time it must not be so strong to generate shear-induced aggregation.\\
The standard procedure to increase colloidal stability is to allow a surfactant or emulsifier to adsorb on the surface of the particles, thus providing with an electrostatic and/or steric stabilization against both Brownian- and shear-induced aggregation.

During particle growth at constant particle number in the course of reaction, the surface of the particles increases and, consequently, a sufficient amount of surfactant must be supplied to make sure that the particles remain covered enough all along the reaction: the fact of having an excessively "naked" surface exposes the particles to a higher number of successful collisions on the hydrophobic polymer spots, which leads to coagulation~\cite{Zaccone_correlation}.
Moreover, the majority of the lab as well as industrial formulations foresee the additions of buffer solutions, such as ammonia or sodium bicarbonate; this is done for various reasons, among which there is pH-control. These additions are especially dangerous for colloidal stability since the added electrolyte species effectively "screen" the electric double-layer (EDL) on the particles surface, thus enhancing the coagulation kinetics.\\
Across the literature, there are plenty of studies focused on the influence of different salts, both mono- and divalents, on the stability of colloidal dispersions, which can quantitatively be described by the Fuchs stability ratio \cite{Spielman}:
\begin{equation}
W = 2 \int_{2}^{\infty}\dfrac{\exp{(U/k_\text{B}T)}}{G(l) l^2} \text{d}l,
\label{Fuchs_Stability_Ratio}
\end{equation}
where $U$ is the interaction potential between two particles, $k_\text{B}$ the Boltzmann constant, $T$ the absolute temperature, $l$ the centre-to-centre distance between the particles normalized by their size, and $G$ a function representing the hydrodynamic lubrication forces between two spherical particles~\cite{Spielman}. \\
The stability ratio $W$ represents the slow down of the aggregation between two particles (with respect to diffusion-limited kinetics) due to the presence of a repulsion barrier caused by the EDL around the particles; the above formula for the stability ratio $W$ has been extended to include the effect of shear flow (which speeds up the kinetics) \cite{Zaccone2009,Zaccone2010,Conchuir2013,Lattuada2016}.
Jia and coworkers \cite{Zichen} introduced a useful approach as they considered the influence of different salts on a carboxylic latex surface, by analysing the impact of association equilibria between counterions and the surface charge groups on $W$: they showed that an increasing number of association events led to a reduced number of active surface charge groups which ultimately caused a reduction of the colloidal stability. By comparing the results from their model to experimental data derived from Static Light Scattering (SLS), they accurately predicted the aforementioned decrease of $W$ with the increase of total salt concentration.

Ehrl et al. \cite{Ehrl2009} have extended this study with the intent of predicting the critical coagulation concentration (CCC) for certain pairs salts/carboxyl-stabilized colloids which is the molar concentration of counterions which causes a colloidal dispersion to instantaneously aggregate. They have predicted the CCC value by finding the concentration of each salt which causes $W$ to be around 1.5 and compared their results to experimental data provided in the literature \cite{Behrens} finding a very good agreement.\\

In spite of these extensive efforts, there are currently no studies in the literature which address the intimate link between the polymerization kinetics, the reaction environment and the coagulation process.
The present work aims to study the colloidal stability of an emulsion polymerization system starting from the beginning of the polymerization reaction. At the same time, since in recent years new updated models for critical processes in the polymerization kinetics, such as radical entry and exit, have been proposed, the following paper has also the aim to include these new state-of-the-art mechanisms into a detailed and predictive kinetic model of the emulsion polymerization process. The resulting framework, calibrated extensively on literature data, allows us to rationalize the coagulation behavior in complex industrial test cases, and provides a quantitative understanding of the subtle interplay between surfactant surface coverage and ionic strength on the colloidal stability across the whole polymerization process.\\

The following work is divided into two parts. The first part (Part one) presents the mathematical model for the emulsion polymerization reaction kinetics, which is described by adapting the \textit{pseudo-homopolymerization} approach.
We then calibrate the model with two different series of test cases, first the homo-polymerization of \textit{n}-butyl acrylate and then the co-polymerizations of \textit{n}-butyl acrylate and methyl methacrylate, from the literature. This analysis will guarantee an accurate description of kinetic variables such as conversion, composition and particle size through a very limited number of adjustable parameters that are determined by comparison with literature data.\\
The rest of the paper (Part two) then focuses on a more complex industrial system, the \textit{n}-butyl acrylate/methyl methacrylate/2-hydroxyethyl methacrylate ter-polymer produced in a 1 $\text{m}^3$ reactor. After having verified once again that the model can reproduce the overall conversion and particle size this time in comparison with novel experimental measurements, two different industrial test cases are analyzed with the model: their formulations foresee different amounts of surfactant with the aim of rationalizing the interplay of different values of surfactant coverage of the particle surface and salt content on the coagulum formation throughout the entire polymerization process.

\section{Modelling of emulsion polymerization kinetics}
According to the established mechanistic picture of the process, an emulsion polymerization  process is divided into three steps \cite{Harkins}. (I) During the first step (Interval I), the particles are formed. The initial dispersion is made of monomer droplets dispersed in an aqueous solution where an initiator is dissolved.. The latter species produces primary radicals which start reacting with the monomers dissolved in water. After having reached a certain degree of polymerization, the oligomers become particle precursors and they are immediately surrounded by the molecules of surfactant which also provides them with electrostatic (if ionic) or steric (if non-ionic) stability against aggregation. Since the monomers are usually hydrophobic, they quickly swell the precursors constituting the so called \textit{particle phase}.
Once the total number of formed particles $N_\text{P}$ becomes constant, the second step (Interval II) where the growth process of the particles takes place: the monomers present in the particle phase are converted into polymer but, at the same time, they are replaced by others which diffuse from the droplets which behave as reservoirs.\\ 
Once the droplet phase has been entirely consumed, the third step (Interval III) begins where the residual monomer within the particle phase is fully depleted.\\
Different modeling approaches have been reported in the literature \cite{Saldivar,Dube,Gao}capable of predicting the following target features
\begin{enumerate}
	\item	Kinetic variables such as conversion and composition;
	\item 	Particle Size Distribution (PSD);
	\item	Molecular Weight Distribution (MWD);
\end{enumerate}
In this project the product is prepared by adopting a seeded polymerization: an already prepared dispersion (seed) causes the process to start directly from step II avoiding particle formation.
Hence, we first followed the methodology proposed by Gao et al. \cite{Gao}:
\begin{enumerate}
	\item The dispersion is considered monodisperse at any time;
	\item The reactor is perfectly mixed;
	\item The overall polymerization rate is equal to the consumption of the monomers in the particle phase, with the consumption of monomer by other reactions (e.g. chain transfer to monomer) as well as in the aqueous phase being negligible.
\end{enumerate}
Let us start the model presentation by introducing the formal definition of the instantaneous conversion $X^\text{inst}$, 
and the overall conversion $X^\text{overall}$ as follows:
\begin{equation}
\begin{cases}
X^\text{inst} = \dfrac{\sum_{j=1}^{\text{N}_m} (m_j^\text{t} - m_j) + m_\text{P,0}}{\sum_{j=1}^{\text{N}_m} m_j^\text{t} + m_\text{P,0}}; \\ \\
X^\text{overall} = \dfrac{\sum_{j=1}^{\text{N}_m} (m_j^\text{t} - m_j) + m_\text{P,0}}{\sum_{j=1}^{\text{N}_m} m_j^\text{tot} + m_\text{P,0}};
\end{cases}
\label{Conversion_Model}
\end{equation} 
where $m_\text{P,0}$ is the initial amount of polymer introduced as seed into at the reactor before the beginning of the monomer additions, $m_j^\text{t}$ is the mass of monomer $j$ added until time $t$, $m_j$ is the unreacted mass of monomer $j$ at time $t$, and the index $j$ runs over the different monomer species. 
In order to evaluate the unreacted mass $m_j$ of the $N_\text{m}$ monomer species, we need to solve the following balances:
\begin{equation}
\dfrac{d m_j}{d t} = \dot{m}_j - M_{\text{p},j}, \quad \text{j} = 1, ... , N_m,
\label{Monomers_Mole_Balance}
\end{equation}
 where the first term on the r.h.s. represents the addition rate (in mass) of component \textit{j}, while $M_\text{P,j}$ is its consumption by the reaction. The latter term is conveniently expressed through the so-called pseudo-homopolymerization approach \cite{Storti1989}, which enables to reduce the evaluation of the overall reaction rate to that of a homopolymer system. Accordingly, the consumption $M_\text{P,j}$ in a homo-polymerization is expressed as
\begin{equation}
	M_\text{p,j} = k_{\text{p},j} [\text{J}]_\text{P} \ \text{MW}_j \ \dfrac{\tilde{n}   N_\text{P}}{N_\text{AV}}
\end{equation}
where $k_\text{p,j}$ is the propagation rate of species $j$, $[\text{J}]_\text{P}$ its concentration within the particle phase, $\text{MW}_j$ its molecular weight, $\tilde{n}$ the average number of radicals per particle, $N_\text{P}$ the total number of particles and $N_\text{AV}$ the Avogadro number. According to the pseudo-homopolymerization approach, the key parameters are in fact the average propagation rate coefficients, $k_\text{p,j}$. Assuming the reactivity of an active chain to be fully determined by its last monomer unit (terminal model), the corresponding average rate constant $\overline{k}_{\text{p},j}$ is expressed as follows:
\begin{equation}
	M_\text{p,j} =  \overline{k}_{\text{p},j} [\text{J}]_\text{P} \dfrac{\tilde{n}   N_\text{p}}{N_\text{AV}} = \biggl( \sum_{i=1}^{N_r} k_{\text{p},ij} P_i  \biggr) [\text{J}]_\text{p} \text{MW}_j \dfrac{\tilde{n}   N_\text{p}}{N_\text{AV}},
\label{Monomer_Consumption_Rate}
\end{equation}
where $k_{\text{p},ij}$ is the propagation rate coefficient between the $i$-th terminal unit of a propagating chain and the $j$-th monomer species,  $P_i$ the probability  of monomer species $i$ of being the last monomer unit of the propagating radical chain, $N_r$ the total number of possible terminal units. Since a seeded system is considered, $N_\text{P}$ is known \textit{a priori}: given the size of the seed particles, $R_\text{P,0}$, such number is given by
\begin{equation}
N_\text{P} = \dfrac{m_\text{P,0}}{\dfrac{4}{3} \pi R_\text{P,0}^3 \rho_\text{P}}.
\end{equation} 
The density of the co-polymer phase $\rho_\text{P}$ is approximated by a weight-averaged value based on the mass fractions $\omega_i$ of each monomer in the solid phase and the densities of their respective homo-polymers $\rho_\text{Pi}$:
\begin{equation}
\rho_\text{P} = \sum_{i=1}^{N_m} \rho_{\text{P},i} \ \omega_i.
\end{equation}
By summing up all of the $M_\text{p,j}$, the mass balance for the growing polymer $m_\text{P}$ can be expressed as:
\begin{equation}
\dfrac{d m_\text{P}}{d t} = \biggl[ \sum_{j=1}^{\text{N}_m} \biggl( \sum_{i=1}^{\text{N}_r} k_{\text{p},ij} P_i \biggr) [\text{J}]_\text{P} \text{MW}_j \biggr] \ \dfrac{\tilde{n} N_\text{P}}{N_\text{AV}}.
\label{Polymer_Mass_Balance}
\end{equation}
Given the overall mass of the particles, that of the single particle $m_1 = m_\text{P}/N_\text{P}$ and, the average particle size $R_\text{P}$ is readily evaluated:
\begin{equation}
R_\text{P}(t) =  \biggl( \dfrac{3}{4 \pi } \ \dfrac{m_1}{\rho_\text{P}}  \biggr)^{1/3}.
\label{Particle_Size_Evaluation}
\end{equation} 
\subsection{Part one: Mechanistic description}
The description of the mechanistic aspects behind the model variables appearing in Eq.(\ref{Monomer_Consumption_Rate}) and Eq.(\ref{Polymer_Mass_Balance}) are introduced in this subsection.
\subsubsection{1) Monomer partitioning}
The concentrations of the monomeric species in each phase, $[\text{J}]_\text{k}$, are needed to evaluate the reaction rates properly. Assuming negligible mass transport resistances, the distribution of each monomer among the three phases is evaluated using partition coefficients $K_j^k $, defined as the ratio of the volume fraction of the $j$-th component between the $k$-th phase (either particle or droplet phase) and the aqueous phase:
\begin{equation}
	K_j^k = \dfrac{\phi_{j,\text{sat}}^k}{\phi_{j,\text{sat}}^w} \sim \dfrac{\phi_{j}^k}{\phi_{j}^w}
\end{equation}
The series of $K_j^k$ coefficients should be specific for the monomers-copolymer system of study. However, since it was not possible to experimentally evaluate them, the partitioning coefficients of each component in its homopolymer have been applied as found in the literature or, as in the case of MMA and 2-HEMA, calculated by knowing the appropriate volume fractions at saturation conditions proposed in Table \ref{Saturation_Volume_Fractions}.\\
\begin{table}
	\centering
	\begin{tabular}{|c|c|}
		\hline   
		\textbf{Property}       	             & \textbf{Value}  \\
		$\phi_{\text{MMA,sat}}^p$\cite{gardonII} & 0.73 \\
		$\phi_{\text{MMA,sat}}^w$ \cite{Maxwell} & 0.027       \\
		$\phi_\text{2-HEMA,sat}^w$\cite{khan2006}& 0.094        \\
		\hline
	\end{tabular}
	\caption{Saturation values adopted to compute the missing partition coefficients}
	\label{Saturation_Volume_Fractions}
\end{table}
The evaluation of the volume fractions $\phi_j^k$, together with the total volume of each phase $V^k$, is carried out solving the algebraic equations reported in the Supporting Information. From these values, the molar concentrations of each monomer are readily evaluated, given their densities $\rho_{\text{m},j}$ and molecular weights $\text{MW}_j$, as follows:
\begin{equation}
[\text{J}]_k = \dfrac{\text{nr moles of the j-th monomer in the k-th phase}}{V^k} = \dfrac{\rho_{\text{m},j} }{\text{MW}_j}
\dfrac{V_j^k }{V^k} = \dfrac{\rho_{\text{m},j}}{\text{MW}_j} \phi_j^k.
\end{equation}
\subsection{2) Active chain end probability}
\begin{figure}
	\centering
	\subfloat{\includegraphics[width = 0.65 \linewidth]{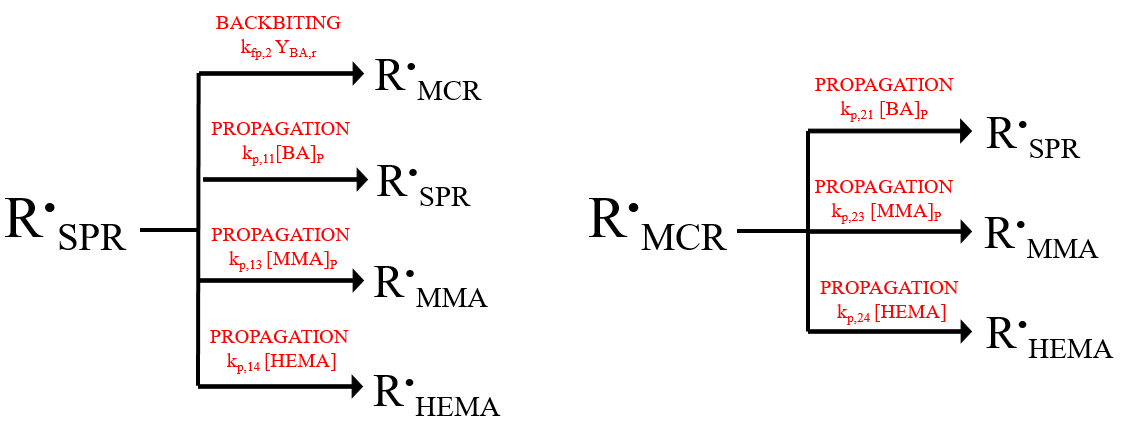}} \quad \quad
	\subfloat{\includegraphics[width = 0.65 \linewidth]{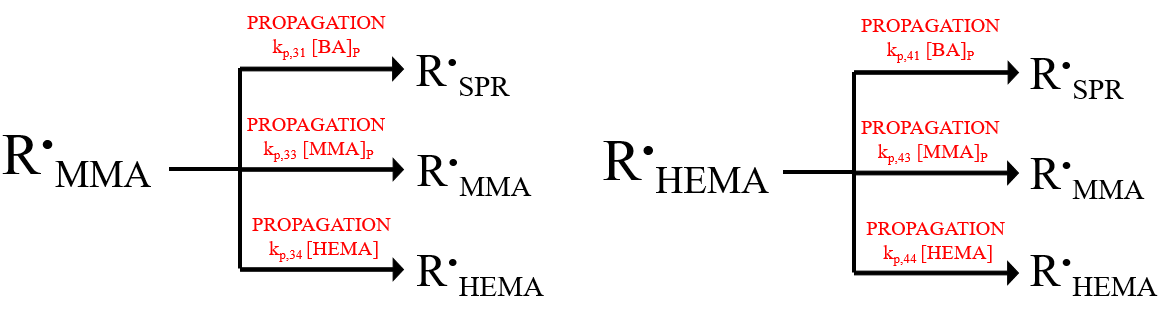}} 
	\caption{Reactions determining active chain type scheme for \textit{n}-BA/MMA/2-HEMA ter-polymerization.}
	\label{fig:Reaction_Scheme}	
\end{figure}

 Each probability $P_i$ is defined as the concentration of chains in the particle phase having a certain "active site" $[\text{R}_i^{\bullet}]_\text{P}$ normalized by their total concentration $[\text{R}_\text{TOT}^{\bullet}]_\text{P}$. Since different monomers as well as different radical types are present, four different $P_i$ values are introduced:
\begin{equation} 
P_1 = \dfrac{[\text{R}_\text{SPR}^{\bullet}]_\text{P}}{[\text{R}_\text{TOT}^{\bullet}]_\text{P}} \quad P_2 = \dfrac{[\text{R}_\text{MCR}^{\bullet}]_\text{P}}{[\text{R}_\text{TOT}^{\bullet}]_\text{P}} \quad P_3 = \dfrac{[\text{R}_\text{MMA}^{\bullet}]_\text{P}}{[\text{R}_\text{TOT}^{\bullet}]_\text{P}} \quad P_4 = \dfrac{[\text{R}_\text{2-HEMA}^{\bullet}]_\text{P}}{[\text{R}_\text{TOT}^{\bullet}]_\text{P}}.
\label{Probabilities_Active_Chains}
\end{equation}
Methyl methacrylate and 2-hydroxyethyl methacrylate allow secondary forms ($\text{R}_\text{MMA}^{\bullet}$ and $\text{R}_\text{2-HEMA}^{\bullet}$) only, whereas \textit{n}-butyl acrylate can assume both secondary ($\text{R}_\text{SPR}^{\bullet}$) and tertiary ($\text{R}_\text{MCR}^{\bullet}$) radical forms; the transition from the first to the second type is an intramolecular chain transfer reaction known as \textit{backbiting} \cite{Li}. Long story short, the transition from one "active site" to the other can happen by "cross-propagation" with a different monomer or by backbiting (\textit{n}-butyl acrylate only).
Here it is important to discuss the effect of the intramolecular chain transfer to polymer on the reaction kinetics: across the literature it has been highlighted that the effect provided by the intramolecular chain transfer to polymer on the polymerization kinetics can be suppressed or, however, provide a very small contribution to the reaction kinetics \cite{hlalele1,hlalele2} according to H-NMR measurements carried on a n-Butyl Acrylate/Styrene copolymer with a minimal mole fraction of styrene of 30\% in the final composition of the latexes.\\ 
For this reason a sensitivity analysis on the impact of the backbiting on the kinetics of the BA/MMA co-polymerization has been conducted and the results have been discussed in Appendix B: it can be demonstrated that the impact of the intramolecular chain transfer to polymer on the kinetic variables is basically negligible when the mole fraction of methyl methacrylate is comparable to the compositions adopted in the aforementioned studies, meanwhile the backbiting still plays an important role when the mole fraction of \textit{n}-BA is 90\% or more. For these reasons it has been decided to consider the effect of the backbiting for every co- and ter-polymerization discussed in this paper.\\
The reactions capable to modify the type of terminal monomer unit of an active chain are schematically shown in Figure \ref{fig:Reaction_Scheme}. 
Given this set of reactions and assuming their dominant role among all the reactions involving active chains, the following balances can be written:
\begin{equation}
\begin{cases}
\dfrac{d P_2}{dt} = 0 = P_1 \biggl( k_\text{fp,2} Y_\text{BA,r} \biggr) - P_2 \biggl( k_\text{p,21} [\text{BA}]_\text{P} + k_\text{p,23}[\text{MMA}]_\text{P} + k_\text{p,24}[\text{2-HEMA}]_\text{P} \biggr); \\ \\ 
\dfrac{d P_3}{dt} = 0 =  P_1 \biggl( k_\text{p,13} [\text{MMA}]_\text{P} \biggr) + P_2 \biggl(k_\text{p,23} [\text{MMA}]_\text{P} \biggr) - P_3 \biggl( k_\text{p,31} [\text{BA}]_\text{P} + k_\text{p,34} [\text{2-HEMA}]_\text{P}\biggr) + \\ + P_4 \biggl( k_\text{p,43} [\text{MMA}]_\text{P} \biggr) ; \\ \\
\dfrac{d P_4}{d t} = 0 = P_1 \biggl( k_\text{p,14} [\text{2-HEMA}]_\text{P} \biggr) + P_2 \biggl( k_\text{p,24} [\text{2-HEMA}]_\text{P} \biggr) + P_3 \biggl( k_\text{p,34} [\text{2-HEMA}]_\text{P}  \biggr)+ \\ - P_4 \biggl( k_\text{p,41} [\text{BA}]_\text{P} + k_\text{p,43}[\text{MMA}]_\text{P} \biggr) \\ \\
P_1 + P_2 + P_3 + P_4 = 1.
\end{cases}
\label{Radical_Balance_2}	
\end{equation}
where the subscripts have been defined as follows: 1 = SPR, 2 = MCR, 3 = MMA, and 4 = 2-HEMA. Moreover, $k_\text{fp,2}$ is the backbiting rate, $k_{\text{p},ij}$ the propagation rate constant between the i-th active site and the j-th monomer, and $Y_{\text{BA},r}$ the mole fraction of reacted \textit{n}-butyl acrylate in the growing polymer:
\begin{equation}
	Y_{\text{BA},r}(t) = \dfrac{m_{\text{BA},r} \ \text{MW}_\text{BA}}{\sum_{i=1}^{N_m} m_\text{i,r}\  \text{MW}_i},
\end{equation}
where $m_\text{i,r}$ are the masses of reacted monomers until time t.
\subsection{3) Average number of radicals}
The average number of radicals per particle is a key factor in determining the rate of consumption of each monomer. Its value is computed by solving the popular Smith-Ewart (SE) equations \cite{SmithEwart}, describing the time evolution of each particle state $N_i$, i.e. the probability of finding a particle containing $i$ propagating chains. The SE equations can be written as
\begin{multline}
\begin{cases}
\dfrac{d N_i}{d t} = \rho N_{i-1} + \overline{k}_{i+1} (i+1) N_{i+1} + c (i+2)(i+1)N_{i+2} - \biggl( \rho + \overline{k}_i i +  c i (i-1)  \biggr) N_i; \\
\tilde{n} = \sum_{i=1}^{\infty} N_i  i,
\end{cases}
\label{Smith_Ewart_General_Equation}
\end{multline}
The first source term of Eq.(\ref{Smith_Ewart_General_Equation}) represent the increase of the number of particles with state $i$ due to entry of a oligomeric radical in a particle with $i-1$ propagating chains with rate $\rho$, meanwhile the second the exit of a monomeric radical from a particle with state $i+1$ with a state-dependent rate equal to $\overline{k}_{i+1}$; finally the third term represent the formation of particle with $i$ active chains due to a termination event in a particle with state $i+2$ happening with rate $c$.
On the other hand, the loss terms in Eq.(\ref{Smith_Ewart_General_Equation}) represent the same phenomena all happening to a particle with state $i$ which subsequently reduce $N_i$ due to the entry of a radical inside it, the exit of a monomeric radical from it or a termination between two chains contained in it; all of the aforementioned rates, $\rho$, $\overline{k}_{i}$ and $c$, will be discussed down below.
This is supposed to be a system of infinite ordinary differential equations (ODEs), so the maximum value of $i$ will be set equal to $i_\text{cr}$, and large enough to ensure that the contribution of states with $ i > i_\text{cr}$ to $\tilde{n}$ is negligible. The SE equations involve several rate coefficients, the evaluation of which is described in detail in the following.
\subsubsection{Termination in the particle phase}
A key quantity in Eq.(\ref{Smith_Ewart_General_Equation}) is $c$, the frequency of bimolecular termination of two propagating chains inside a particle:
\begin{equation}
c = \dfrac{\langle k_\text{t} \rangle_\text{p}}{2 N_\text{AV} V_s} 
\label{Termination_Rate_Particle_Phase_1}
\end{equation}
$\langle k_\text{t} \rangle_\text{p}$ is the rate constant of bimolecular termination in the particle phase. In order to find it it is first necessary to introduce its equivalent in bulk conditions, when the monomer conversion is close to zero (negligible polymer concentration). According to the same procedure used for the average propagation rates $k_{\text{p},j}$ introduced in Eq.(\ref{Monomer_Consumption_Rate}), it is defined as: 
\begin{equation}
	\langle k_\text{t} \rangle_\text{b} = \sum_{i=1}^{N_r} \sum_{j=1}^{N_r} k_{\text{t},ij} P_i P_j,
\label{Termination_Rate_Bulk}
\end{equation}
where $k_\text{t,ij}$ is the termination rate constant under bulk conditions between two propagating chains having an i-th and a j-th active site, respectively; for now the chain length dependence on the termination rate has been neglected.
The termination rate in the particle phase will be then formalized as:
\begin{equation}
    \langle k_\text{t} \rangle_\text{p} = \langle k_\text{t} \rangle_\text{b} \exp{\biggl[- \biggl( \sum_{i=1}^{N_m} Y_i a_i\biggr)\biggr]}
    \label{Termination_Rate_Particle_Phase_2}
\end{equation}
The exponential term in Eq.(\ref{Termination_Rate_Particle_Phase_2}) represents the slow down of the termination in the particle phase with respect to its zero conversion value by the so called gel effect. This effect mainly depends on the amount of growing polymer in the particle phase $\phi_\text{pol}^P$ and on its composition: different amounts of converted monomers provide with distinct influences on the gel effect through monomer-specific coefficients $a_i$ weighted by their respective mole fractions $Y_i$ in the polymer
\begin{equation}
	Y_i =  \dfrac{(m_{\text{i},r} + m_{\text{i},s}) \ \text{MW}_i}{\sum_{j=1}^{N_m}(m_{\text{j},r} + m_{\text{i},s})\ \text{MW}_j},
	\label{Overall_Mole_Fraction_BA}
\end{equation}
where $m_{\text{i},s}$ are the residual masses of each monomer in the initial seed.
\subsubsection{Exit rates $\overline{\textbf{k}}_\textbf{i}$}
At this point we introduce a novel approach to describe the remaining contributions to the SE equations starting from the second term on r.h.s. in Eq.(\ref{Smith_Ewart_General_Equation}), the loss term. This term contains the state-dependent desorption rate constant of monomeric radicals $k_i$ resulting from chain transfer to monomer~\cite{Ghielmi}. Since we are dealing with multiple monomeric species we will describe $\overline{k}_i$ as a superposition of the exit rates of every possible monomer involved: 
\begin{equation}
\overline{k}_i = \sum_{j=1}^{N_m} k_i^\text{j}
\end{equation}
Note that the desorption of 2-HEMA has been neglected because we didn't find any reliable value of the corresponding chain transfer to monomer rate coefficient. However, the impact of this assumption on the model predictions is expected to be minimal, given the low amount of this monomer in the system.\\
The complete description of the exit rates has been proposed in the Supporting Information according to the state-dependent approach adopted by Ghielmi et. al:\cite{Ghielmi}
\begin{equation}
k_i^j = \biggl( \sum_{k \in N_r'} k_\text{fm,kj} P_k [\text{J}]_\text{p} \biggr) Q_i^j \biggl( 1 + (1-\beta_j) \dfrac{N_{i-1}}{i N_i} \dfrac{\sum_{i=1}^\infty i Q_i^j N_i}{1- (1-\beta_j) \sum_{i=1}^\infty Q_i^j N_{i-1}}\biggr).
\end{equation}
$N_r'$ is the number of radical species a certain monomer $m$ can have: as mentioned earlier \textit{n}-BA can assume both secondary and tertiary radical forms, so $N_r' = 1,2$, while $N_r' = 3$ for MMA. The first big brackets represent the formation of radical monomers due to chain transfer, while the second term in the last big brackets takes into account the probability that a desorbed radical could re-enter a particle with state $N_{i-1}$.
In these equations $Q_i^j$ expresses the probability for a monomeric radical to desorb from a particle in the i-th state rather than propagating with another monomer or terminate with another chain:
\begin{equation}
Q_i^j = \dfrac{k_{\text{dm},j}}{k_{\text{dm},j} + R_\text{p}^j + 2 c^j (i-1)},
\end{equation}
In particular, each propagation rate $R_\text{p}^j$ can be written as 
\begin{equation}
R_\text{p}^j  = \sum_{k \in N_r'} \sum_{l=1}^{N_m} (k_{\text{p},kl} P_k [\text{l}]_\text{P}),
\end{equation}
while the respective termination rates $c^j$ in the particle phase is expressed as
\begin{equation}
c^j  = \dfrac{\sum_{k \in N_r'} P_k ( \sum_{l=1}^{N_r} k_{\text{t},kl} P_l )}{2 N_\text{AV} V_S} \exp{\biggl[ - \biggl(\sum_{i=1}^{N_m} a_i Y_i \biggr)\phi_\text{pol}^P \biggr] }
\end{equation}
while $\beta_j$ is the probability for a desorbed radical to react in the aqueous phase (terminate or propagate) instead of re-entering another particle. In this scenario we have assumed complete re-entry as the ultimate fate of the desorbed radicals, an hypothesis already adopted in the homo-polymerizations of \textit{n}-Butyl Acrylate \cite{Maxwell_BA} and Methyl Methacrylate\cite{Ballard}, which means $\beta_j = 0$ for both the radicals.\\
Finally, $k_\text{dm,i}$ is the classic Smoluchowski diffusion-limited rate model which considers the relative amounts of each monomer in the particles through their volume fractions $\phi_i^\text{p}$ and their respective diffusion coefficients in water $D_\text{w}^j$ \cite{Gilbert_Review}:
\begin{equation}
k_\text{dm,j} = \dfrac{3 D_\text{w}^j}{R_S^2} \dfrac{ [\text{J}]_\text{w}}{[\text{J}]_\text{P}},
\end{equation} 
where $R_{S}$ is the radius of the swollen particle (i.e. the volume of the $P$  phase divided by the number of particles), a quantity which is updated at every step during the simulation of the polymerization process.
\subsubsection{Entry rates}
On the other hand, the source term of Eq.(\ref{Smith_Ewart_General_Equation}), i.e. the first term on the r.h.s., is represented by the entry rate of new radical species ($\rho$) into each particle, which is given by two different contributions:
\begin{equation}
\rho = \rho_I + \rho_\text{re}.
\end{equation}
Here, $\rho_I$ is the contribution associated to the oligomeric radicals produced in aqueous phase by decomposition of the initiator, while $\rho_\text{re}$ is the re-entry rate of monomer radicals desorbed by other particles, to be defined below.\\ 
The mechanism of radical entry into the particle is well known, as its rate determining step (rds) has been studied extensively. In the past it  was generally thought that the diffusion of oligomers (diffusion model) rather than their collision with particles could represent most of the process, but it has been demonstrated that the entry is basically independent from any events happening on the particle surface. More recent works have established that the rds is probably the propagation in the aqueous phase\cite{Maxwell}; the proposed mechanism has been named "control by aqueous phase growth" and it has been confirmed by the independence of the entry rate from the particle size.
We have decided to describe $\rho_\text{I}$ by adopting its steady state approximation~\cite{Maxwell} 
\begin{equation}
\rho_\text{I} = \dfrac{2 f k_\text{d} \text{I} N_\text{AV}}{N_\text{p}} \biggl( \dfrac{2 \sqrt{f k_\text{d} [\text{I}]_\text{w} \langle k_\text{t,w} \rangle}}{\sum_{i=1}^{N_m} \overline{k}_{\text{p},i}^\text{w} [\text{i}]_\text{w}}  + 1 \biggr)^{1-\overline{z}}
\label{Entry_Rate_Maxwell_Morrison}
\end{equation}
where $k_\text{d}$ is the decomposition rate of the initiator, $f$ is its efficiency and I is the number of moles of persulphate within the liquid phase and $\langle k_\text{t,w} \rangle$ is the average termination coefficient in the aqueous phase defined as a geometric average of the homotermination rates of the involved monomers:
\begin{equation}
\langle k_\text{t,w} \rangle = \biggl( \prod_{i=1}^{N_m} k_\text{t,w}^i \biggr)^{1/N_m}.
\end{equation}
While the values of $k_\text{t,w}^i$ are readily available for MMA and HEMA, two different types of radical have to be considered for n-BA, secondary and tertiary. The corresponding rate constants $k_\text{t,11}$ and $k_\text{t,22}$ differ by two orders of magnitude\cite{BarthBA}; therefore, only the rate constant $k_\text{t,w}^\text{BA}$ is an unknown parameter, for which a value has been chosen in between those reported for the secondary and the tertiary radical types.\\
$\sum_{i=1}^{N_m} \overline{k}_{\text{p},i}^\text{w} [i]_\text{w}$ is the total propagation rate in aqueous phase and the associated rate constants $\overline{k}_{\text{p},i}^\text{w}$ related to the consumption of each i-th monomer and, finally, $\overline{z}$ is the average degree of polymerization of the oligomers entering the particles. It has been evaluated as an average of the critical degrees of polymerization of the respective homopolymers $z_i$.
The complete explanation behind the evaluation of Eq.(\ref{Entry_Rate_Maxwell_Morrison}) is proposed in the Supporting Info.\\
On the other hand, $\rho_\text{re}$ is the re-entry rate of monomeric radicals previously desorbed written as a function of the average number of radicals per particle in the system $\tilde{n}$, and of the average radical desorption constant which is the superposition of the average desorption rates of the single monomers:
\begin{equation}
\begin{cases}
\langle k_i^j \rangle = \biggl( \sum_{k \in N_r'} k_\text{fm,kj} P_k [\text{J}]_\text{p} \biggr) \dfrac{\sum_{i=1}^{\infty}i Q_i^j N_i}{\tilde{n} \biggl( 1- (1-\beta_j) \sum_{i=1}^{\infty} Q_i^j N_{i-1} \biggr)}; \\ 
\rho_\text{re} = \biggl( \sum_{m}\langle k_i^j \rangle (1-\beta_j) \biggr) \tilde{n} .
\end{cases}
\end{equation}
With this calculation scheme, we will be able to numerically solve the SE Eqs.(\ref{Smith_Ewart_General_Equation}) to obtain $\tilde{n}$, which can then be used as input to compute the time-evolution of particle size by means of Eqs.(\ref{Polymer_Mass_Balance})-(\ref{Particle_Size_Evaluation}).
The average particle size as a function of time in the polymerization process, $R_\text{P}(t)$, calculated in this way, will then serve as input to quantify the surfactant coverage of the particles at all times during the process, as described in the following sections.

\subsection{Part two: Effects of surfactant and salt content}
The main focus of the second part of this work is the study of the influence of different particle surface coverage by surfactant and salt contents on the colloidal stability of the ter-polymerization system.\\
The prediction of the particle size via Eq.(\ref{Particle_Size_Evaluation}), using $\tilde{n}$ from the solution of the SE scheme Eqs. (\ref{Smith_Ewart_General_Equation}), provides the input to evaluate the particle surface covered by surfactant molecules. To this aim, one has to implement a mole balance which describes the partitioning of the surfactant between the particle and the aqueous phase, under the assumption that the adsorption of the surfactant on droplets is negligible. Thus the mass balance reads as:
\begin{equation}
S(t) = A_\text{P} \Gamma + S^\text{w}(t),
\label{Surfactant_Split}
\end{equation}
where $ A_\text{P} = 4 \pi R_\text{P}^2 N_P$ is the total surface of the particles, with $R_P$ evaluated through Eq.(\ref{Particle_Size_Evaluation}), $\Gamma$ is the concentration of emulsifier adsorbed over a single particle, and $S^\text{w}$ is the number of moles of surfactant in the aqueous phase.\\
To model $\Gamma$ as a function of the molar concentration of the surfactant in the aqueous phase $[S]_\text{w}$ we have adopted a two-step Langmuir adsorption model \cite{ZacconeJPCB}:
\begin{equation}
\Gamma =   \Gamma_\infty \dfrac{ k_1 [S]_\text{w}(n^{-1} + k_2 [S]_\text{w}^{n-1})}{1 + k_1  [S]_\text{w}(1 +  k_2 [S]_\text{w}^{n-1})},
\label{two_step_adsorption}
\end{equation}
where $k_1$ represents the adsorption of single surfactant molecules on the surface of the particles, while $k_2$ represents the formation rate of hemimicelles with aggregation number $n$.\\
Finally, $\Gamma_\infty$ is the concentraton of absorbed emulsifier at saturation which is related to the area occupied by a single molecule of surfactant $a_s$ by
\begin{equation}
\Gamma_{\infty} = \dfrac{1}{N_{\text{AV}} \ a_s}
\end{equation}
Unfortunately, there is no availability of experimental data for the particular system of interest, so we have taken input data from the literature on a quite similar system, i.e. an acrylate co-polymer stabilized by stearate ionic surfactant\cite{ZacconeJPCB}.\\
At the same time, we shall keep track of the total salt content inside the system because the presence of counterions inside the liquid medium screens the negative surface charges provided by the surfactant, which can lead to an overall loss of colloidal stability of the dispersion.
According to the formulation, all the compounds that have been used are ammonium and sodium persulphates together with ammonia and a carboxylate salt of potassium, so their chemical dissociation equilibria can be written as
\begin{equation}
\begin{cases}
	\text{K Carb} \rightarrow \text{K}^+ + \text{Carb}^{-} \\
	\text{Na}_2 \text{S}_2\text{O}_8  \rightarrow 2 \text{Na}^+ +  \text{S}_2 \text{O}_8^{2-} \\
	(\text{NH}_4)_2 \text{S}_2\text{O}_8  \rightarrow 2 (\text{NH}_4)^+ +  \text{S}_2 \text{O}_8^{2-} \\
	\text{NH}_3 + \text{H}_2 \text{O}   \rightleftharpoons \text{NH}_4^+ + \text{OH}^{-}
\end{cases}
\label{Dissociation}
\end{equation}
We have considered all the persulphates and the carboxylate to be strong salts, while a weak dissociation for the ammonia is appropriate:
\begin{equation}
	\gamma = \dfrac{[\text{NH}_4]_\text{w}^+ [\text{OH}^-]_\text{w}}{[\text{NH}_3]_\text{w}}.
\end{equation}
The monitored variable will be the overall amount of the counterions in the aqueous phase during the polymerization:
\begin{equation}
	[\text{CI}] = [\text{Na}^+]_\text{w} + [\text{NH}_4^+]_\text{w} + [\text{K}^+]_\text{w}
\end{equation}
\section{Input Data}
The values of all of the kinetic parameters needed for the validation of the test cases are presented in the following. The propagation, termination and transfer to monomer rate constants, since the process have temperatures which vary in the range 343 K - 351 K, are defined according to the Arrhenius form:
\begin{equation}
k = \text{A} \exp{ \biggl( -\dfrac{E_\text{a} [\text{kJ} \text{mol}^{-1}] }{R T} \biggr)}.
\end{equation}
\begin{table}
	\centering
	\begin{tabular}{|c|c|c|}
		\hline 
		\textbf{Rate} & \textbf{A} $[\textbf{l}/(\textbf{mol} \ \textbf{s})] $& $\textbf{E}_\textbf{a} [\textbf{kJ/mol}]$  \\ 
		$k_\text{d}$\cite{Borisov}             & 3.08$\cdot 10^{13}$   & 118.0   \\
		$k_\text{p,11}$ \cite{Plessis}         & 2.05$\cdot 10^7$      & 17.89   \\
		$k_\text{p,21}$ \cite{BarthBA}         & 9.20$\cdot 10^5$      & 28.30  \\
		$k_\text{p,33}$ \cite{Beuermann}       & 2.67$\cdot 10^6$      & 22.36 \\
		$k_\text{p,44}$ \cite{achilias}        & 8.89$\cdot 10^6$      & 21.89 \\
		$k_\text{t,11}$\cite{BarthBA}  & 1.30$\cdot 10^{10}$   & 8.40  \\
		$k_\text{t,22}$\cite{BarthBA}  & 9.00$\cdot 10^6$      & 5.60   \\
		$k_\text{t,12}$\cite{BarthBA}  & 4.20$\cdot 10^9$      & 6.60   \\
		$k_\text{t,33}$\cite{BarthMMA}            & 2.33$\cdot 10^{10}$   & 8.44  \\
		$k_\text{t,44}$\cite{achilias}  & 3.91$\cdot 10^7$      & 5.26  \\
		$k_\text{fm,11}$ \cite{Maeder}    & 0.016 $k_\text{p,11} $  & 15.2  \\ 
		$k_\text{fm,21}$ \cite{Maeder}    & 0.016 $k_\text{p,22}$    & 15.2 \\
		$k_\text{fm,32}$ \cite{Sangster}  & 2.00$\cdot 10^5$      & 46.10  \\
		$k_\text{fm,31}^*$         & 0              &  0    \\
		$k_\text{fm,32}^*$         & 0              &  0    \\
		$k_\text{fm,12}^*$         & 0              &  0    \\
		$k_\text{fp,2}$\cite{BarthBA}          & 1.6 $\cdot 10^8$     & 34.7 \\
		\hline
	\end{tabular}
	
	\caption{Arrhenius parameters for propagation, termination and transfer rates; $k_{\text{fm},ij}^*$: value assumed in this work.}
	
	\label{Kinetic_Parameters_Test_Cases}
\end{table}
Concerning the cross-propagation rates $k_{\text{p},ij}$, these have been evaluated upon adopting the respective reactivity ratios
\begin{equation} 
k_{\text{p},ij} = \dfrac{k_\text{p,ii}}{r_{ij}},
\label{Hetero-Propagation-Rates}
\end{equation}
while the cross-termination ones $k_\text{t,ij}$, in lieu of experimentally measured input which is not available, have been evaluated as a geometric average
\begin{equation}
k_{\text{t},ij} = \sqrt{k_{\text{t},ii} k_{\text{t},jj}}.
\label{Hetero-Termination-Rates}
\end{equation}
In Table \ref{Extra_Inpout_Data} the values of the remaining input parameters, the ones independent of temperature, are shown.
\begin{table}
	\centering
	\begin{tabular}{|c|c|c|}
		\hline   
		\textbf{Property}       	   & \textbf{Value} & \textbf{Source}   \\
		$f$\cite{Asua_BA}              &        0.6     &    Literature      \\
		$K_\text{BA}^p$\cite{Gugliotta}&       480      &  Literature        \\
		$K_\text{2-HEMA}^p\cite{chu2000}$	   &       3.51     &  Literature\\
		$K_\text{MMA}^p$        	   &       27       &  Table\ref{Saturation_Volume_Fractions} \\
		$K_\text{BA}^d$\cite{Gugliotta} &       740      &  Literature        \\
		$K_\text{2-HEMA}^d$     	   &       10.70    & Table\ref{Saturation_Volume_Fractions} \\
		$K_\text{MMA}^d$        	   &       36       &  Table\ref{Saturation_Volume_Fractions}\\
		$z_\text{BA}$           	   &       3        &  Eq.(\ref{Homopolymerization_DOPs}) \\
		$z_\text{2-HEMA}$       	   &       31       &  Eq.(\ref{Homopolymerization_DOPs}) \\
		$z_\text{MMA}$          	   &       6        &  Eq.(\ref{Homopolymerization_DOPs}) \\
		$r_\text{13}$  \cite{Asua_BA_MMA} &       0.414    &  Literature \\
		$r_\text{31}$ \cite{Asua_BA_MMA}  &       2.24     &  Literature \\
		$r_\text{14}$   \cite{varma1976}	  &       0.167    & Literature \\
		$r_\text{41}$ \cite{varma1976}        &       5.404    & Literature \\
		$r_\text{34}$ \cite{varma1976}        &       0.284    & Literature \\
		$r_\text{43}$  \cite{varma1976}       &       1.016    & Literature \\
		$D_\text{w}^\text{BA}  \ [\text{m}^2 \ \text{s}^{-1}]$  & 1.5 $\cdot 10^{-10}$  &    \\
		$D_\text{w}^\text{MMA} \ [\text{m}^2 \ \text{s}^{-1}]$  & 2.42 $\cdot 10^{-10}$ &    \\
		$n$\cite{ZacconeJPCB}                    &     5.73        &   Literature \\
		$k_1$  [$\text{L} \ \text{mol}^{-1}$]\cite{ZacconeJPCB}  & 1.44 $\cdot 10^{5}$   &   Literature \\
		$k_2$  [$(\text{L} \ \text{mol}^{-1})^{1-n}$]\cite{ZacconeJPCB}  		  & 1.6 $\cdot 10^{20}$   &   Literature\\
		$a_s$ [$\text{m}^{2}$]\cite{ZacconeJPCB}                         		  & 26 $\cdot 10^{-20}$   &   Literature \\
		$\gamma \ [\text{mol} \ \text{L}^{-1}]$                & 1.88 $\cdot 10^{-5}$  &     \\
		\hline
	\end{tabular}
	\caption{Additional non-Arrhenius parameters.}
	\label{Extra_Inpout_Data}
\end{table}
\section{Model parameters identification using literature data}
\begin{table}
	\centering
	\begin{tabular}{|c|c|}
		\hline      
		\textbf{Variable} & \textbf{Value} \\
		\hline
		$a_\text{BA}$& 6.0    \\
		$a_\text{MMA}$ & 16  \\
		$k_\text{t,w}^\text{BA} \ [\text{L} \text{(mol s)}^{-1}] $ & 3.8$\cdot 10^{6}$ \\
		\hline
		
	\end{tabular}\\	
	\caption{Values of the unknown parameters which are determined from the calibration of the mathematical model on test cases from the literature. }
	\label{Adjustable_Parameters}
\end{table}

\begin{figure}
	\centering
	\includegraphics[width = 0.6 \linewidth]{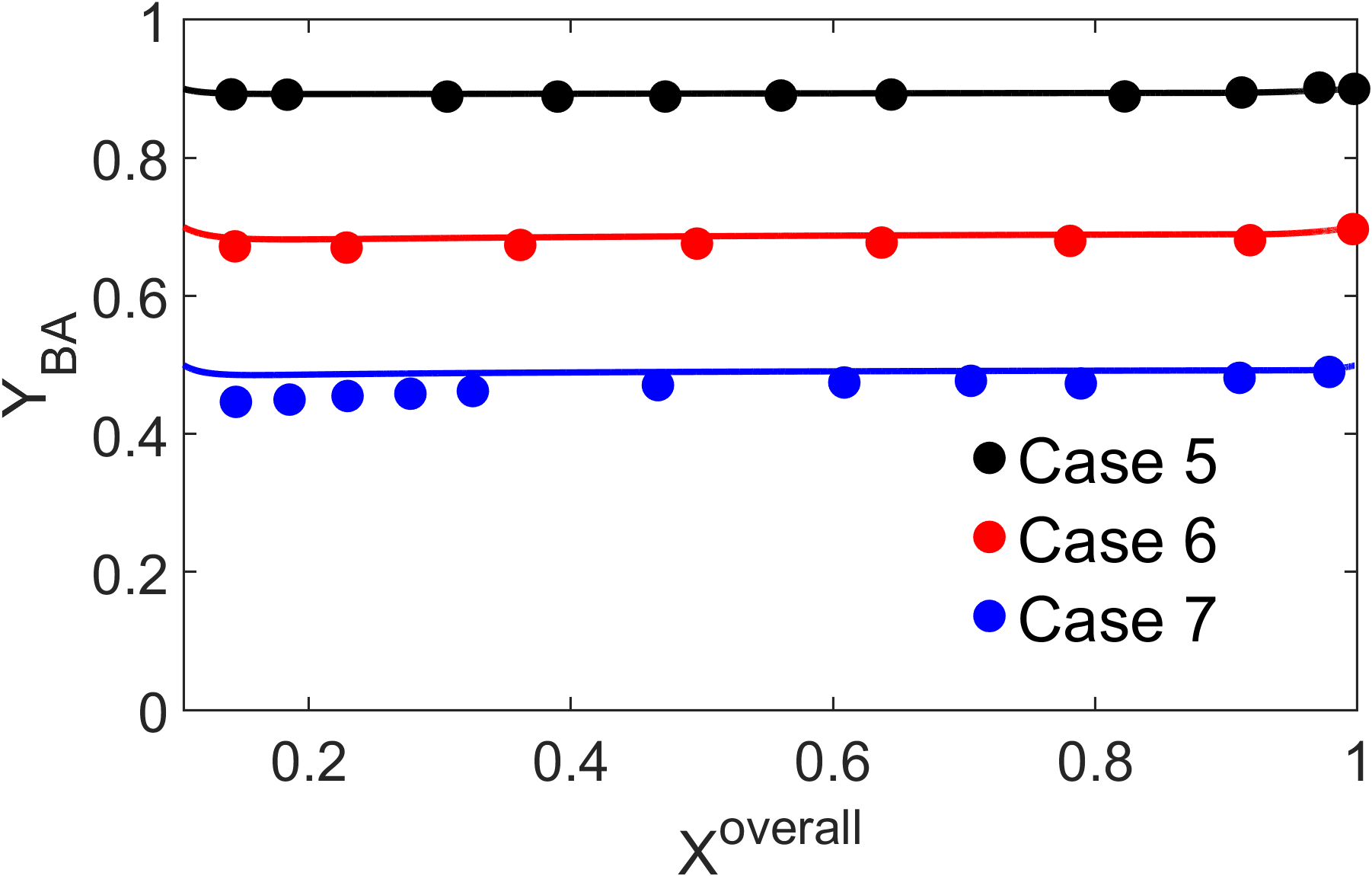}
	\caption{Evolution of the molar fractions of \textit{n}-butyl acrylate in the different runs introduced in Table \ref{Table_Case_Study_Co_Polymerization} compared to experimental data \cite{Asua_BA_MMA}.}
	\label{Composition_BA_MMA}
\end{figure}

In order to determine the unknown parameters, we calibrate the mathematical model by comparing its predictions with literature data of instantaneous and overall conversion, $X^\text{inst}$ and $X^\text{overall}$, for a series of polymerizations of \textit{n}-butyl acrylate and a \textit{n}-butyl acrylate/methyl methacrylate co-polymer. The procedure is based on tuning the monomer-specific gel effect coefficients $a_i$, together with the homotermination rate constant of \textit{n}-BA in water $k_\text{w}^\text{t}$: we will first tune the values to the \textit{n}-BA during its homo-polymerization.. Then, the estimated values have been used as input parameters for the second test case to find the gel effect coefficient for MMA. The estimated parameter values are summarized in Table
 \ref{Adjustable_Parameters}.\\
\begin{table}
	\centering
	\begin{tabular}{|c|c|c|c|c|}
		\hline      
		\textbf{Case} & \textbf{Seed size} [m] & $\textbf{m}_\textbf{p,0} \ [\text{Kg}]$ & \textbf{ \% Initiator}\textsuperscript{\emph{a}} & \textbf{Feed time} [min] \\
		\hline
		1  & 43.01$\cdot 10^{-9}$ & 0.020 & 0.3 & 60  \\
		2  & s. v. & s. v. & 0.3 & 120 \\
		3  & s. v. & s. v. & 0.3 & 180 \\
		4  & s. v. & s. v. & 0.3 & 240 \\
		\hline
	\end{tabular}\\
	
	\textsuperscript{\emph{a}} with respect to total monomer content;
	
	\caption{Formulations for the homo-polymerization validation; additional information is available in the original paper \cite{Asua_BA}.}
	\label{Table_Case_Study_Homo_Polymerization}
\end{table}
\begin{figure}
	\centering
	\includegraphics[width = 0.6 \linewidth]{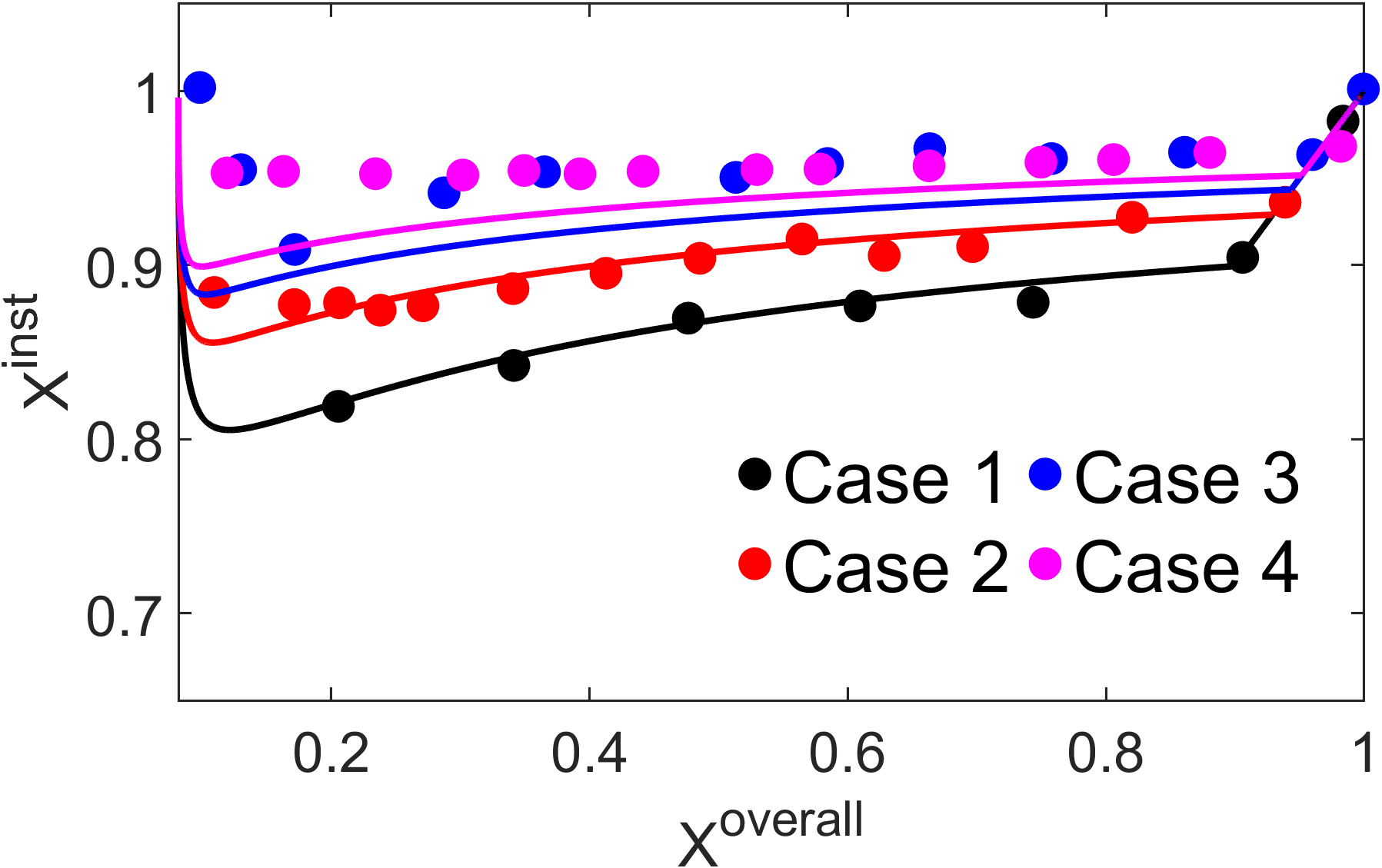}
	\caption{Comparison of $X^\text{inst}$ vs $X^\text{overall}$ obtained by the model (solid lines) and the respective experimental data (symbols) related to the homo-polymerization of \textit{n}-BA adopting different feeding times; for every case the initiator percentage is 0.3\% with respect to the total monomer content. Legend: Case 1 - 1h; Case 2 - 2h ; Case 3 - 3h ; Case 4 - 4h.}
	\label{Model_Validation_BA}
\end{figure}
The formulations for the homo-polymerizations of $\textit{n}$-Butyl Acrylate are presented in Table \ref{Table_Case_Study_Homo_Polymerization} and the comparisons between the experimental data and the model predictions are shown in Figure  \ref{Model_Validation_BA}.
The procedure starts from the homo-polymerization of \textit{n}-Butyl Acrylate considering experimental trends of $X^\text{inst}$ vs $X^\text{overall}$ found in the literature.\\
First, a brief discussion behind the choice of the termination rate in the aqueous phase for \textit{n}-BA $k_\text{t,w}^\text{BA} = 3.8 \ 10^6 \ [\text{L} \ \text{mol}^{-1} \text{s}^{-1}]$ is needed.
According to the Arrhenius parameters provided in Table \ref{Extra_Inpout_Data} the termination rates for \textit{n}-BA secondary and tertiary radicals are $k_\text{t,11} = 7.14 \ 10^{8} \  [\text{L} \ \text{mol}^{-1} \text{s}^{-1}]$ and $k_\text{t,22} = 1.26 \ 10^{6} \  [\text{L} \ \text{mol}^{-1} \text{s}^{-1}]$, respectively at T = 348 K.
The decision to select a value much closer to $k_\text{t,22}$ derives from an analysis of the relative presence of secondary and tertiary radicalic species with respect to the reaction temperature. It has been verified experimentally that at temperatures way below zero, near 213 K, almost all of the radicals assumes the secondary form, meanwhile at temperatures close to the ones of the formulations the vast majority of the radicals are tertiary \cite{Willemse}. For this reason, it has been decided to fix a value for $k_\text{t,w}^\text{BA}$ with the same order of magnitude as $k_\text{t,22}$.\\
Next, the \textit{n}-BA-specific gel effect coefficient is tuned in order for the model to correctly describe the trends $X^\text{inst}$ vs $X^\text{overall}$ for Case 1 to 4; this procedure ends up with a final value of $a_\text{BA} = 6$.
First, in Figure  \ref{Model_Validation_BA} the comparison between the model predictions and the experimental data related to test cases Case 1 to Case 4, where the difference among the formulations is the feeding time of the delayed additions, is proposed.
Even if the model slightly underestimates the data at lower conversions for Case  3 and Case  4 an overall satisfactory match is reached for every test case where the increase of $X^\text{inst}$ at equal $X^\text{overall}$ and considering longer feeding times is correctly reproduced, evidence that the model can reliably predict the amount of monomer converted at any time during the reaction.\\
\begin{figure}
	\centering
	\subfloat{\includegraphics[width = 0.46 \linewidth]{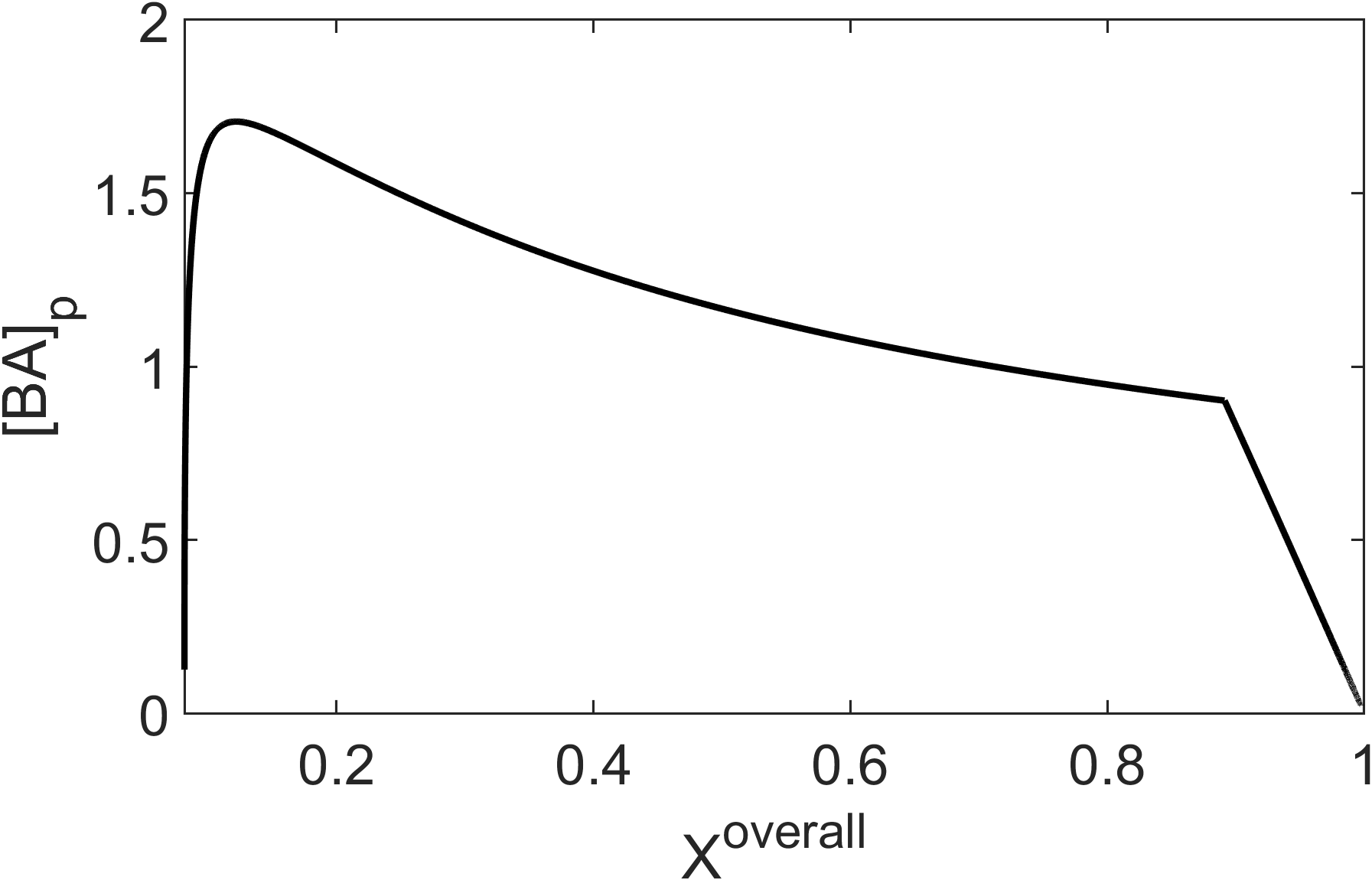}} \quad
	\subfloat{\includegraphics[width = 0.443 \linewidth]{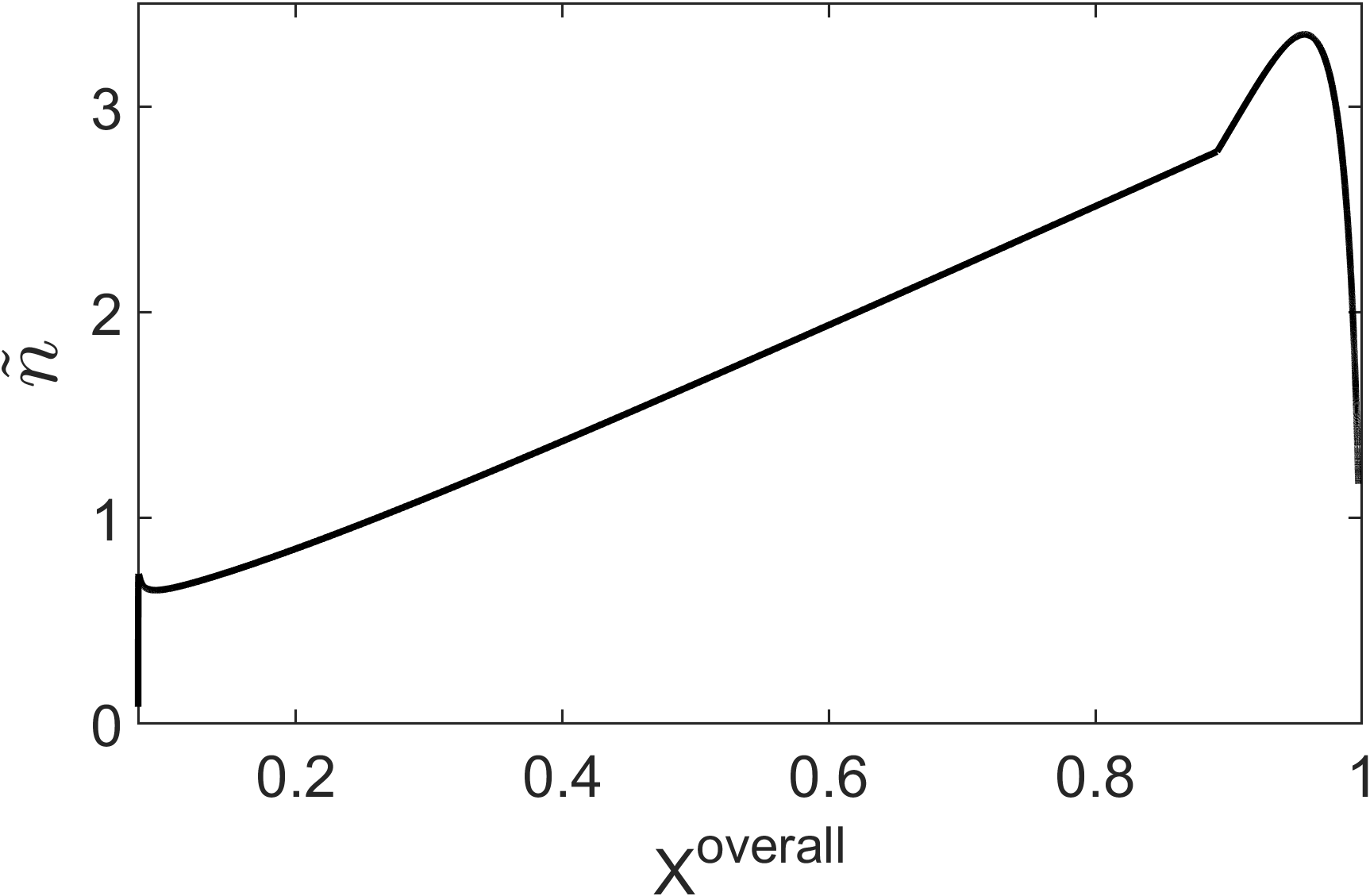}}
	\caption{Concentration of monomer swollen in the particles (left) and average number of radicals (right) for Case  1.}
	\label{BA_1_Kinetic_Data}
\end{figure} 
Nonetheless, Figure  \ref{Model_Validation_BA} proposes a non-trivial trend for the instantaneous conversion which needs further explanations; the discussion will be focused on Case 1, but this explanation can be extended to all the homo- and co-polymerization test cases. The instantaneous conversion $X^\text{inst}$ is influenced by two opposed contributions: on the one hand $X^\text{inst}$ is inversely proportional to the constant feeding rate because higher rates  increase $m_\text{BA}^t$, the total mass of monomer added up until a certain time $t$, which causes the instantaneous conversion to decrease according to its formal definition in Eq.(\ref{Conversion_Model}). On the other hand $X^\text{inst}$ is directly proportional to the radical activity, represented by the average number of radicals $\tilde{n}$,  which increases the consumption of monomer $M_\text{p,BA}$ according to Eq.(\ref{Monomer_Consumption_Rate}).\\
At the beginning the amount of monomer provided is higher than its consumption, so $X^\text{inst}$ starts decreasing, a feature confirmed by the increasing of the amount of \textit{n}-Butyl Acrylate swollen into the particle phase proposed on the left hand side of Figure  \ref{BA_1_Kinetic_Data}. In the meanwhile, the consistent increase of radical activity proposed on the right hand side of the same figure causes $M_\text{p,BA}$ to increase up until it becomes equal to the feeding rate of the monomer, which corresponds to the minimum of $X^\text{inst}$, and subsequently higher causing $X^\text{inst}$ to progressively increase which is also confirmed by the decrease of the concentration of monomer in the particle phase; in the end the discontinuity in the trend of Figure~\ref{Model_Validation_BA} corresponds to the end of the delayed additions from which, by definition, $X^\text{inst} = X^\text{overall}$.\\
\begin{table}
	\centering
	\begin{tabular}{|c|c|c|c|c|c|}
		\hline      
		\textbf{Case} & \textbf{Seed Size} [m]& $\textbf{m}_\textbf{p,0} \ [\text{Kg}]$  & $\textbf{Y}_\textbf{BA,s}$\textsuperscript{\emph{a}} & \textbf{ \% Initiator}\textsuperscript{\emph{b}} & \textbf{Feed time} [min] \\
		\hline
		5 & 42$\cdot 10^{-9}$  & 0.063 & 0.9 & 0.37  & 180  \\
		6 & 44.5$\cdot 10^{-9}$& s.v.  & 0.7 & 0.37   & 180  \\
		7 & 41$\cdot 10^{-9}$  & s.v.  & 0.5 & 0.185  & 180  \\
		\hline
		
	\end{tabular}\\
	
	\textsuperscript{\emph{a}} Mole fraction of \textit{n}-butyl acrylate in the seed equal to the one in the feed; \\
	\textsuperscript{\emph{b}} With respect to the total amount of monomer added.
	\caption{Formulations of \textit{n}-BA/MMA co-polymers; additional information has been reported by Elizalde et al. \cite{Asua_BA_MMA}.}
	\label{Table_Case_Study_Co_Polymerization}
\end{table}
\begin{figure}
	\centering
	\includegraphics[width = 0.6 \linewidth]{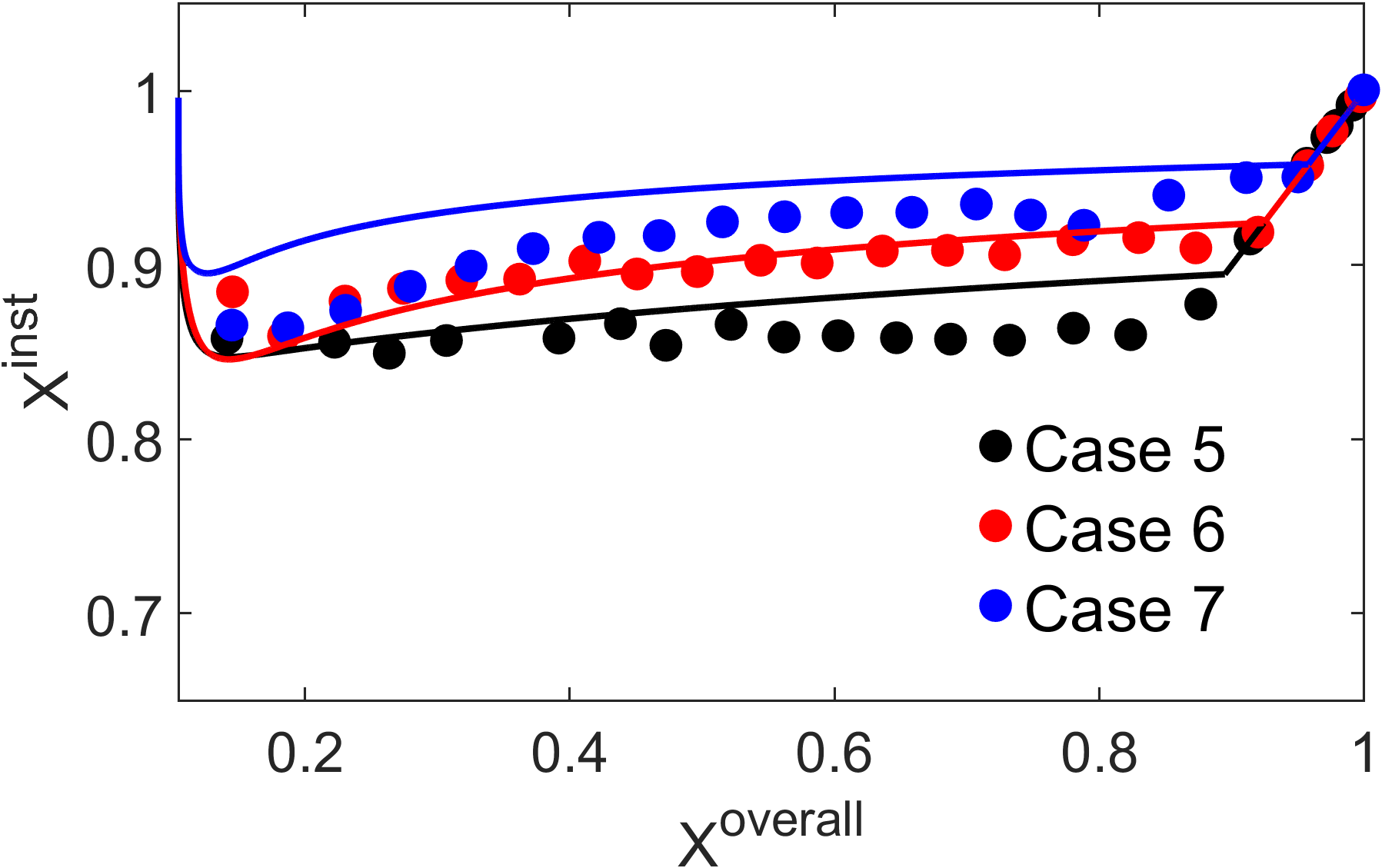}
	\caption{Comparison of $X^\text{inst}$ vs $X^\text{overall}$ obtained from the model and the respective experimental data related to the co-polymerization of \textit{n}-BA/MMA adopting a feeding time of 3h with different feed compositions. Legend: Case 5 - $\text{Y}_\text{BA}$ = 0.9; Case 6 - $\text{Y}_\text{BA}$ = 0.7 ; Case 7 - $\text{Y}_\text{BA}$ = 0.5. Additional information  about the formulation has been reported by Elizalde et al \cite{Asua_BA_MMA}.}
	\label{Model_Validation_BA_MMA}
\end{figure}
Next, we focus on the series of \textit{n}-BA and MMA co-polymerizations whose formulations are proposed in Table \ref{Table_Case_Study_Co_Polymerization}. These experimental data are compared to the model predictions in Figure  \ref{Model_Validation_BA_MMA}. Also in this case the agreement is quite good both qualitatively and quantitatively for different feed compositions and amounts of initiators added, even if the description of Case 7 slightly overestimates $X^\text{inst}$ during the early stages.\\
For the same binary system, the evolution of the copolymer composition with conversion is shown in Figure \ref{Composition_BA_MMA}, the composition as cumulative mole fraction of \textit{n}-BA calculated by Eq.(\ref{Overall_Mole_Fraction_BA}):the overall agreement is indeed satisfactory.\\
To conclude, the proposed mathematical model describes quite accurately crucial kinetic variables such as monomers conversion and product composition during a seeded emulsion polymerization under different conditions.

\section{Industrial Test Case}
Next we consider the more complex case of the \textit{n}-butyl acrylate/methyl methacrylate/2-hydroxyethyl methacrylate ter-polymer. The missing fitting parameter, the gel coefficient for 2-HEMA, is first estimated by comparing the model predictions to the overall conversion measured by Gas Chromatography (GC) and the particle size experimentally evaluated by Dynamic Light scattering (DLS).
The second part of the section presents the analysis of the surfactant surface coverage and salt content for two different runs of the product based on a filtration (grit) analysis carried out on different samples taken during the polymerization process. This procedure allows us to study the interplay between the surfactant and the salt content effects, and to see whether or not the coagulation formation can be predicted based on the actual formulations of the polymerization process.
For proprietary reasons we can not disclose all of the details of the industrial formulation, but relevant information can be found in Table \ref{Pre_Emulsified_Addition}. One should note that the difference in the percentages between Case 8 and Case 9 led to an increase of 18\% on the mass of total surfactant added.
\begin{table}
	\centering
	\begin{tabular}{|c|c|c|c|c|c|c|}
		\hline      
		\textbf{Case} & $\textbf{Y}_\textbf{BA,s}$ & $\textbf{Y}_\textbf{BA,f}$ & \textbf{Feed time} [min] & \textbf{ \% Initiator} \textsuperscript{\emph{a}} & \textbf{Surfactant \%} \textsuperscript{\emph{a}} & \textbf{Ammonia \%} \textsuperscript{\emph{a}} \\
		\hline
		8 & $\sim$ 0.5 & $\sim$ 0.9   & 270 &   0.3  &  0.8   &  0.683  \\
		9 & $\sim$ 0.5 & $\sim$ 0.9   & 270 &   0.3  &  0.97   &  0.683  \\ 
		\hline
	\end{tabular}\\
	
	\caption{Information about the seed and the pre-emulsified addition. $Y_\text{BA,s}$ is the mole fraction of \textit{n}-BA in the seed and $Y_\text{BA,f}$ its equivalent in the feed.}
	\textsuperscript{\emph{a}} With respect to the total mass of monomers;
	\label{Pre_Emulsified_Addition}
\end{table}
\subsection{Experimental section}
Since specific experimental techniques have been applied in the ternary system, the materials and the procedures of the synthesis the analysis of the ter-polyme are summarized below.

\subsubsection{Materials}
\textit{n}-butyl acrylate (Arkema), methyl methacrylate (Dow) and 2-hydroxyEthyl methacrylate (Dow) have been used as monomers, sodium persulphate (Univar), ammonium persulphate (Univar) have been used as initiators in combination with carboxylate salt of potassium (Synthomer LTD) as surfactant and ammonia (Univar) as the buffer solution; every material has been used as received. We can not disclose the exact composition of the surfactant for proprietary reasons. All industrial runs have been conducted using de-ionized water.

\subsubsection{Polymerization process}
The experimental tests involving the industrial test cases (Case 8 and Case 9) were carried out in a 1 $\text{m}^3$ mechanically stirred reactor. First, the temperature of the reactor is raised to 350 K by an external coil where steam is injected. The seed (solids content = 30\%, particle size of the seed can not be disclosed for proprietary reasons) has then been loaded followed by a shot of sodium persulphate. Immediately thereafter, a stream of pre-emulsified monomers together with the surfactant and a second stream containing ammonium persulphate have been slowly added for about 4.5 hours. This interval of time will be referred to subsequently as "\textit{feed additions}". During the feed additions,  the reactor was cooled down by flushing water in the coil in order to maintain the temperature at the desired value. The rotational speed of the impeller is increased during this interval from 40 to 60 rpm to guarantee a proper mixing, which is hindered during the polymerization by the increase of the solids content. Moreover, a series of shots of ammonia has been added during the feed addition in order to maintain the pH as alkaline as possible and facilitate post-processing treatments. At the end of the polymerization, a final shot of buffer solution is added after 90 minutes from the end of the feed additions.
\subsubsection{Characterization of the colloidal samples}
A series of samples has been collected once every 30 minutes during the feed additions and a final one at the end of the polymerization.\\
The amount of free monomer contained in every sample has been measured by \textit{Gas Chromatography} using a Shimadzu 2010 gas chromatograph and AOC 6000 auto sampler together with a FID detector.
From this procedure we are capable of evaluating the concentration of free monomer [FM] (expressed in PPM) inside the sample which, under the condition of perfect mixing, we have associated to the experimental overall conversion $X^{\text{overall}}_{\text{exp}}$ as:
\begin{equation}
	X^{\text{overall}}_{\text{exp}} = \dfrac{(\sum_{i=1}^{N_m} m_i^t) - [\text{FM}] \ 10^{-6} \ m_\text{system}}{\sum_{i=1}^{N_m} m_i^\text{tot}},
\end{equation}
where $m_\text{system}$ is the overall mass contained inside the reactor.\\
The average particle sizes have been evaluated through \textit{Dynamic Light Scattering} (DLS) using a Malvern ZetaSizer Since the solid content in the samples is really high (up to 50\%) we can not run the measurements immediately after sampling, but large dilution is required to perform the DLS successfully; the whole procedure is described in the supporting info.\\
Finally, concerning the \textit{filtration analysis}, we have gathered from each sample a certain mass $m_\text{s}$ which has been filtered through a disposable sieve with empty mass $m_0$ and a cutoff size of 45 $\mu$m. Afterwards the sieves were dried at 80$^{\circ}$ for 10 minutes to eliminate any remaining liquid. Finally, the sieves were weighted a second time to evaluate $m_\text{f}$ and calculate the amount of coagulated colloid, expressed in PPM, as:
\begin{equation}
\text{PPM}\biggl( \dfrac{\text{mg}}{\text{Kg}} \biggr) = \dfrac{m_\text{f} - m_\text{0}}{m_\text{sample}} \ 10^6
\end{equation}
\subsection{Results and discussion}
The first step is the tuning of the influence of the gel effect provided by 2-HEMA on the termination within the particle phase again by finding the best fit for the overall conversion evaluated by GC for Case 9: from the procedure it results that $a_\text{2-HEMA} = 75$.
\begin{figure}
	\centering
	\subfloat{\includegraphics[width = 0.45  \linewidth]{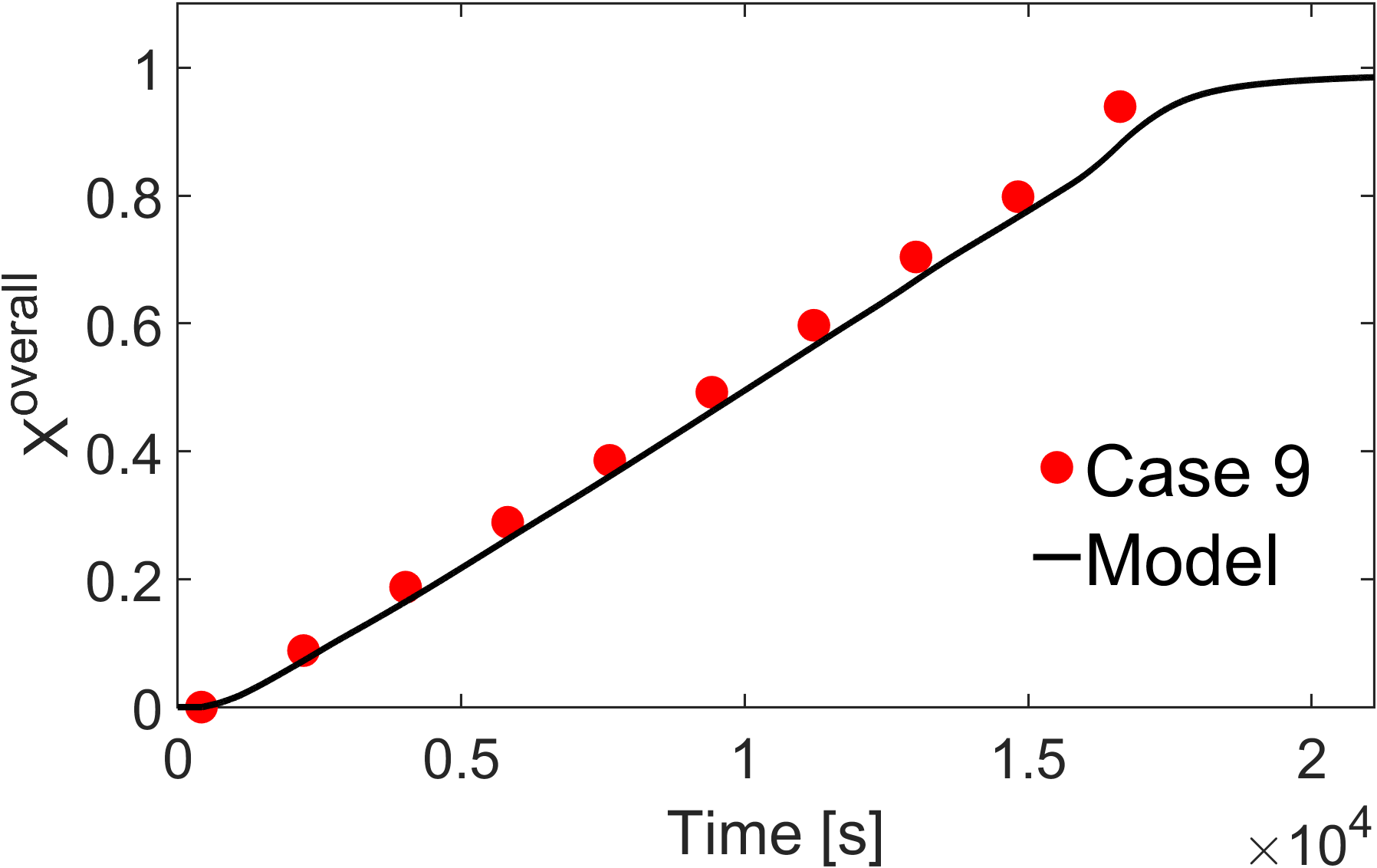}} \quad
	\subfloat{\includegraphics[width = 0.45  \linewidth]{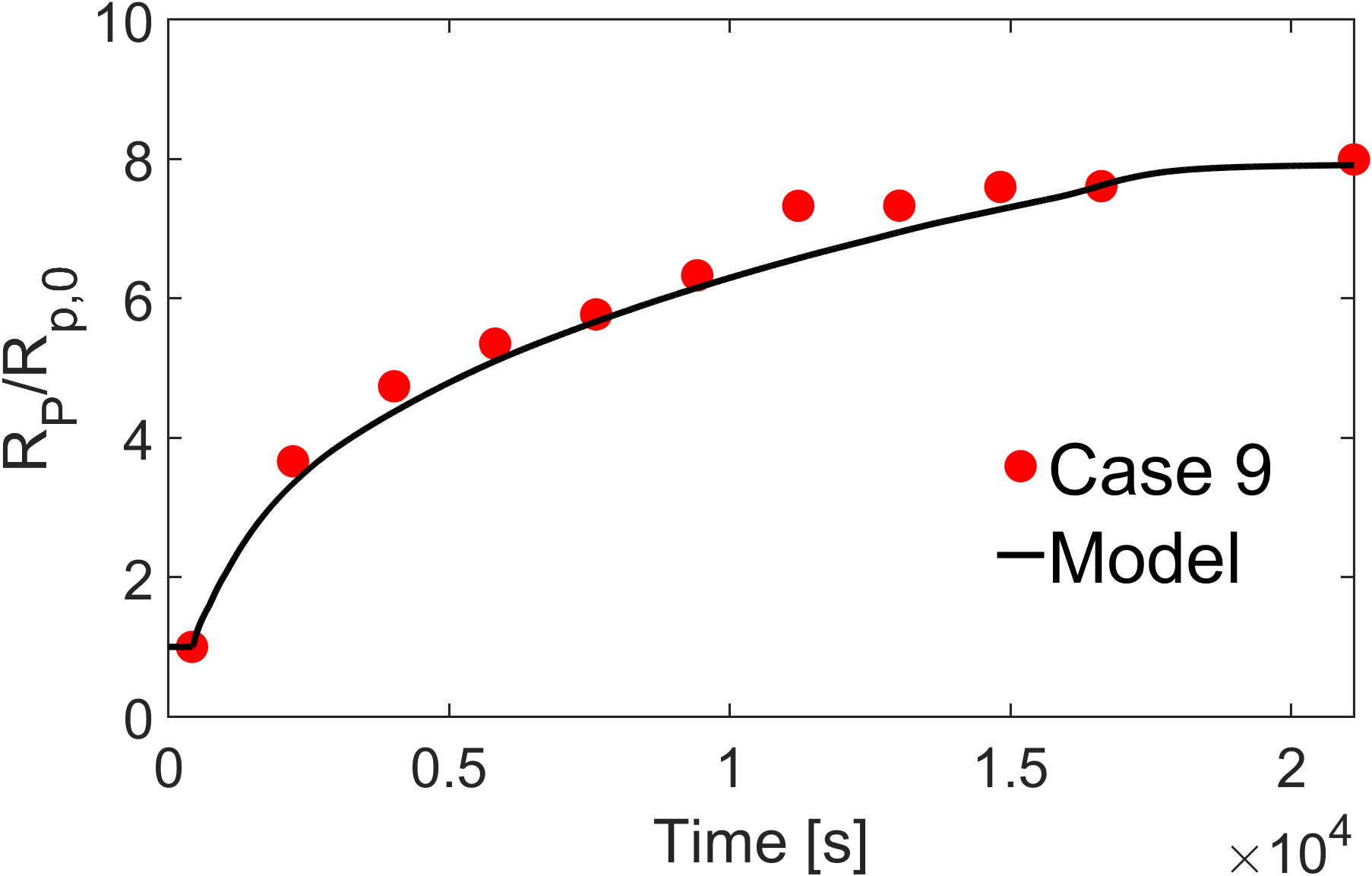}}
	\caption{Temporal trend of the overall conversion and normalized average particle size during the ter-polymerization predicted by the model compared to experimental data.}
	\label{BN21303_kinetic_variables}
\end{figure}
In Figure \ref{BN21303_kinetic_variables} the final prediction of $X^\text{overall}$ and of the normalized average particle size $R_\text{P}/R_\text{P,0}$ ($R_\text{P,0}$ is the radius of the initial seed) are shown, in comparison with the experimental data from the GC and DLS measurements, respectively. Even if there is a slight underestimation of the conversion values, the description of the kinetic variables through the model is satisfactory and it has finally been proved that the model can also reproduce accurately the average particle size of the latex of interest.
\begin{figure}
	\centering
	\includegraphics[width = 0.5 \linewidth]{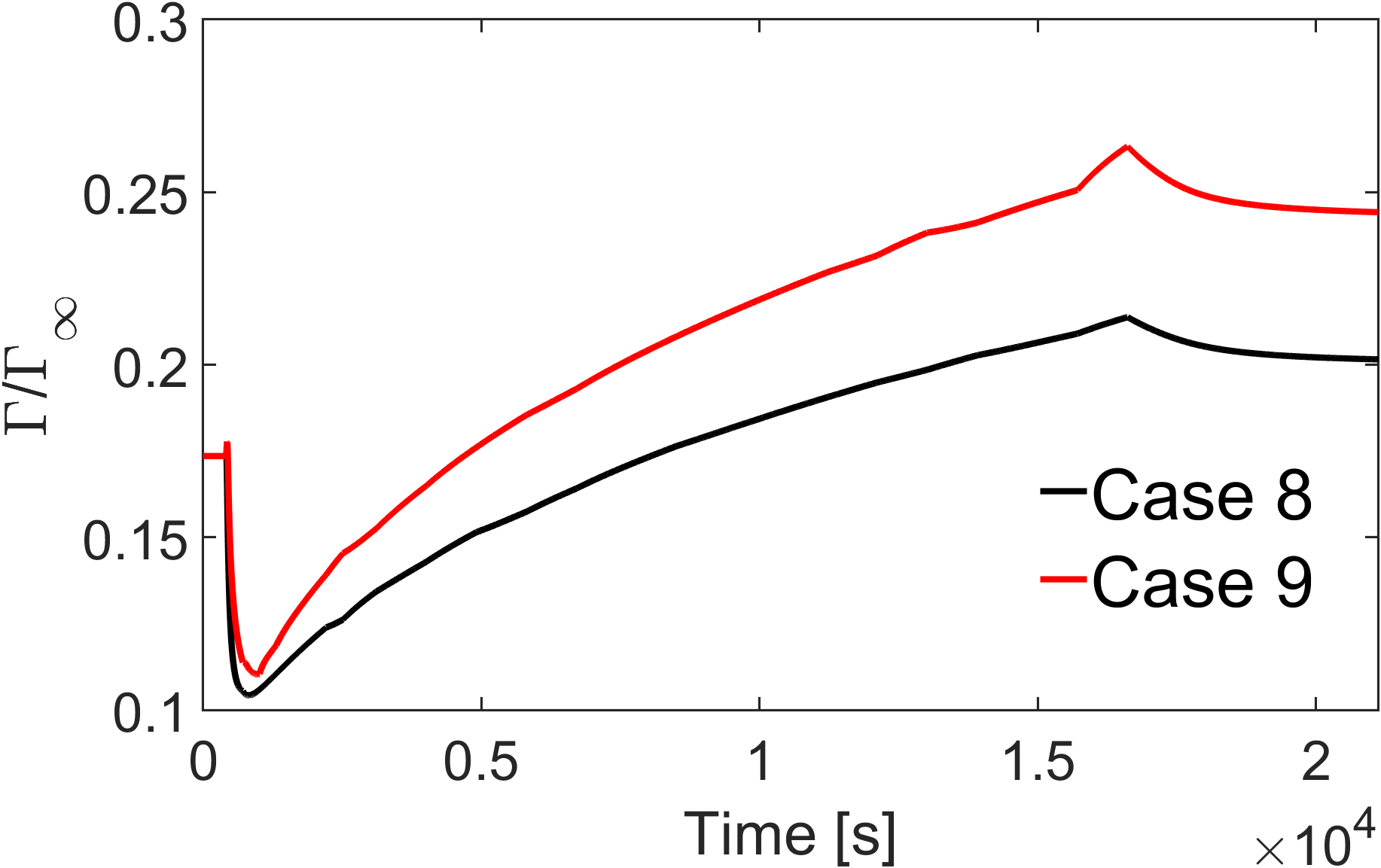}
	\caption{Surface coverage difference between two different batches with different surfactant content.}
	\label{Surface_Coverage}
\end{figure}
\begin{figure}
	\centering
	\subfloat{\includegraphics[width = 0.45 \linewidth]{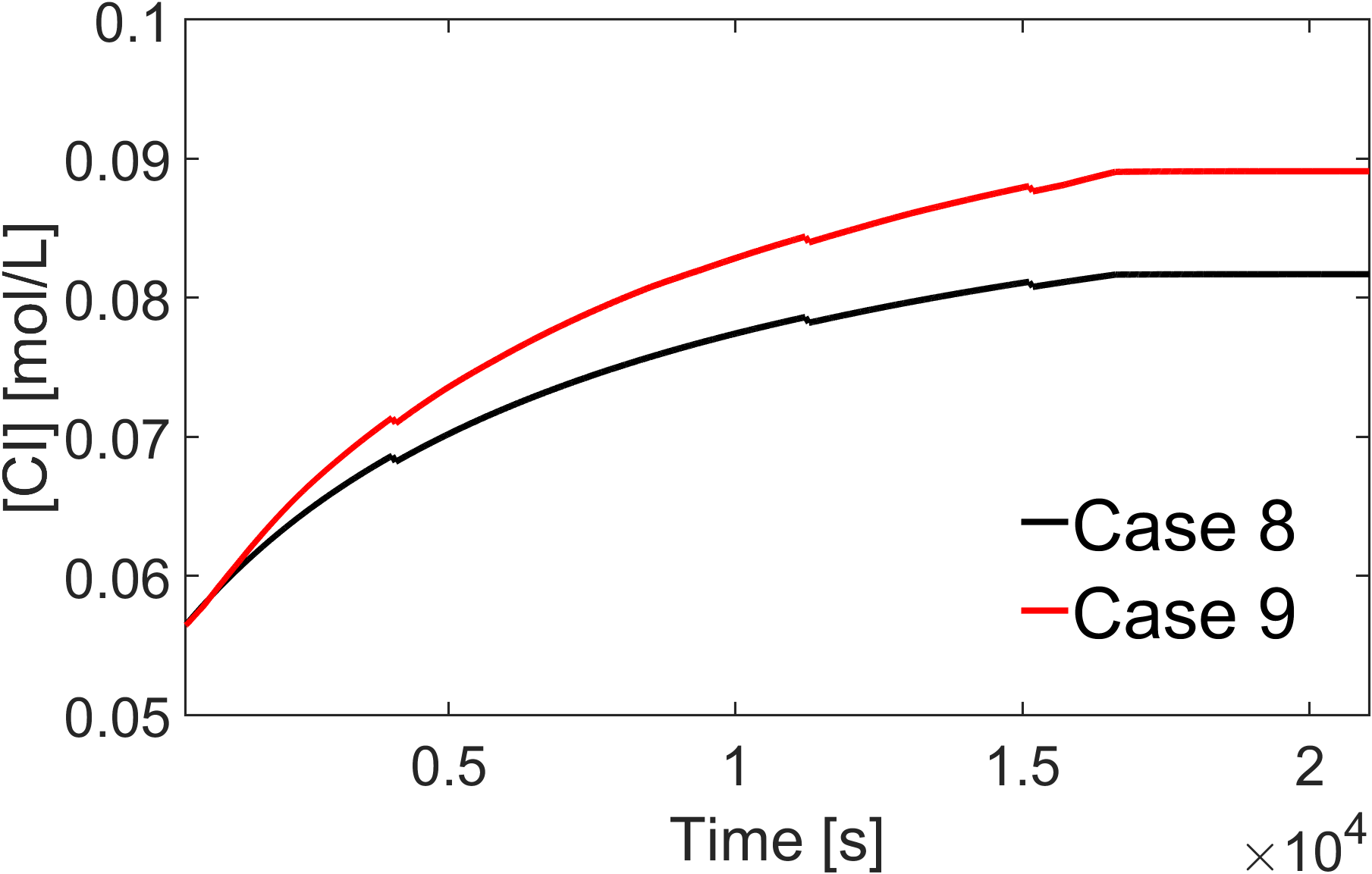}} \quad
	\subfloat{\includegraphics[width = 0.45 \linewidth]{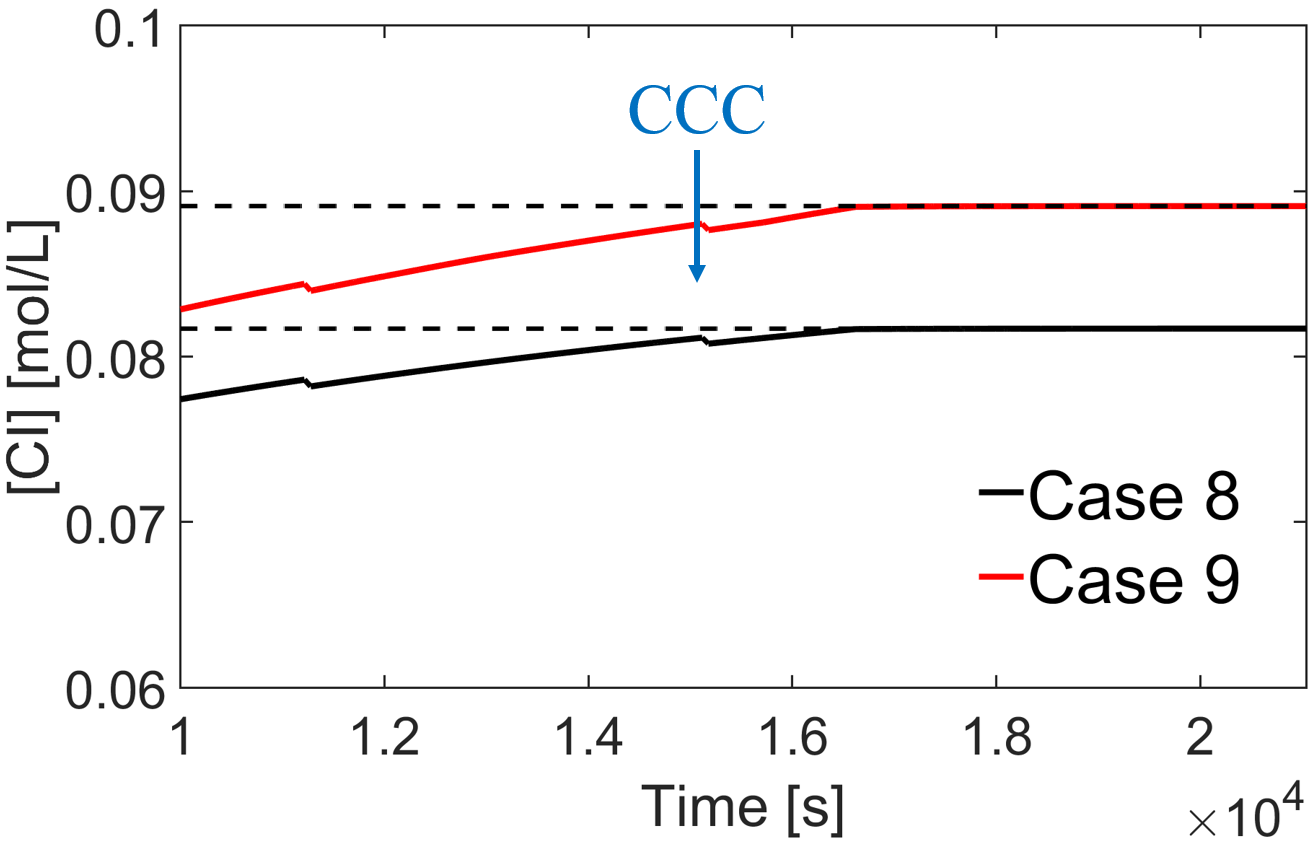}}
	\caption{Concentrations of counterions in the aqueous phase as a function of time during the polymerization process. The panel on the right is a zoomed in version of the plot on the left hand side focusing on the late stage of the process. The dashed lines indicate a lower and an upper bound for the CCC in the system, such that $\text{CCC} \approx 0.085-0.09$ [mol/L].}
	\label{Cationic_Concentration}
\end{figure}
Next, taking advantage of the predicted values of average particles size the total surface of the colloidal dispersion has been evaluated in order to estimate and the average surfactant surface coverage as a function of time through Eq.(\ref{two_step_adsorption}) for both Case 8 and Case 9. This information is shown in Figure \ref{Surface_Coverage}: the initial surface coverage is not null due to a small amount of surfactant derived from the formulation of the seed. At the beginning a decrease is observed  because the total surface area increases with a really high rate for the first 6-7 minutes of the feeding time. Once the growth of $A_\textsuperscript{P}$ starts slowing down then the surface coverage starts increasing as expected until the end of the feed additions where no more surfactant is provided, but at the same time the particles keep growing by depleting the residual monomers.\\
Now, the 20\% extra surfactant provided in Case 9 leads to a higher surface covered by surfactant: a final gap of 20\% at the end of the feed additions is reached.\\
On the other hand, in Figure \ref{Cationic_Concentration}, we show the difference in salt content due to the additional surfactant amount added in Case  9. We recall that the total counterion concentration [CI] includes the contribution of every monovalent counterion in the system, which are [$\text{Na}^+$], [$\text{NH}_4^+$] and [$\text{K}^+$]: the difference keeps increasing until a maximum value of 7\% is reached at the end of the feed additions which remains constant until the end of the process. This relatively small gap, as we will see below, leads to a dramatically different coagulation behavior.
\subsection{Filtration analysis of the coagulation behavior}
\begin{figure}
	\centering
	\subfloat{\includegraphics[width = 0.4625 \linewidth]{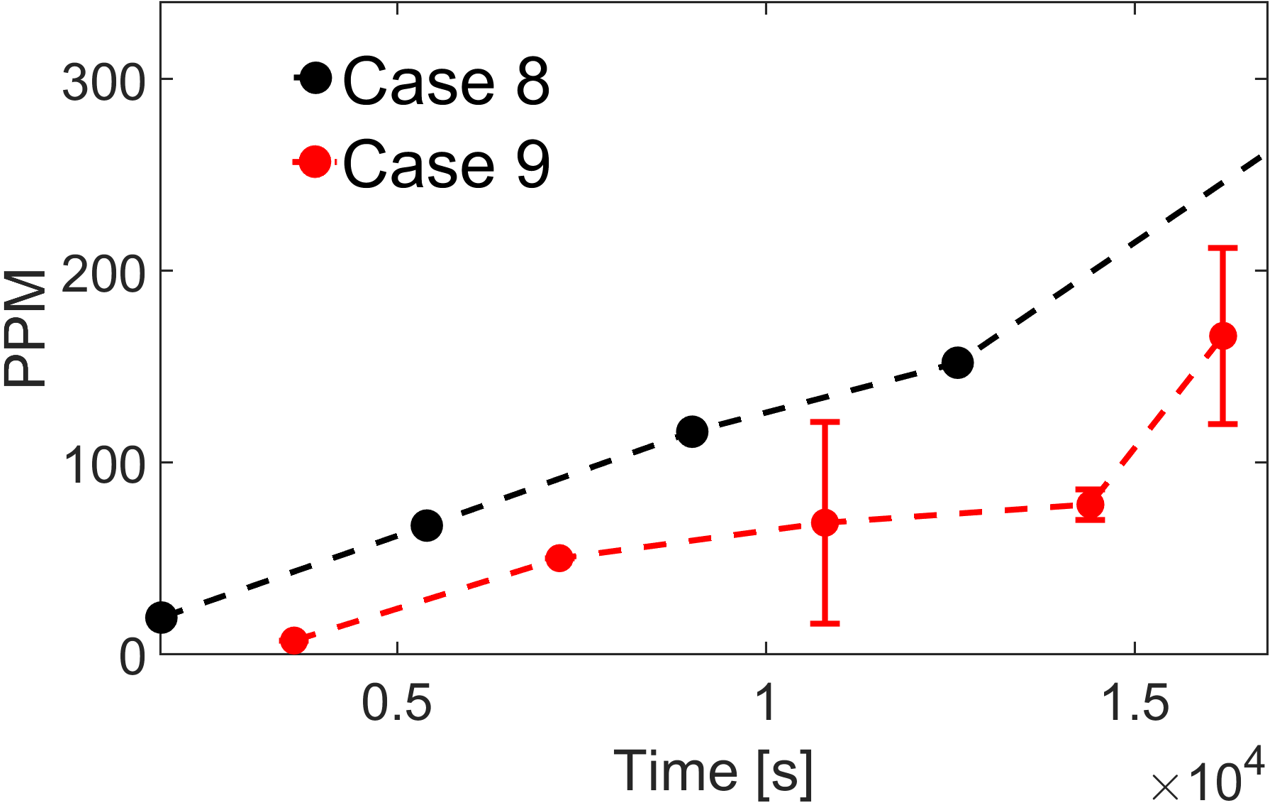}} \quad
	\subfloat{\includegraphics[width = 0.48 \linewidth]{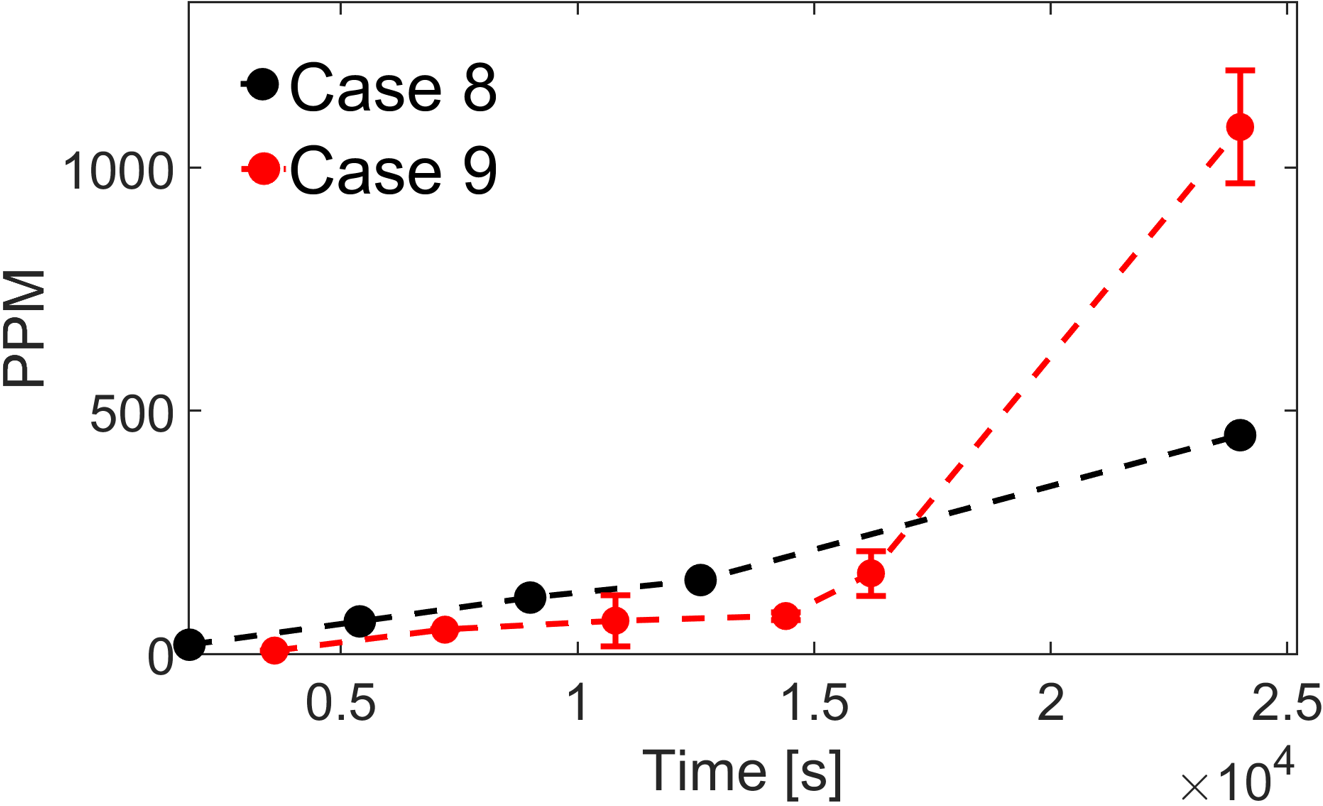}}
	\caption{Temporal evolution of the coagulation behavior from the grit analysis, in the two different batches: on the left the plot is focused on the time-span of the feed additions, on the right the behavior during the whole polymerization process is shown.}
	\label{Grit_Analysis}
\end{figure}
The final point of our analysis aims at understanding the link between the amount of colloidal coagulum detected via the  characterization procedure described above, and the predicted values of surfactant surface coverage and salt content.\\
It can be seen on the left hand side of Figure \ref{Grit_Analysis} that, during the feed additions, the the stabilizing effect due to the increase of surfactant surface coverage is predominant because the PPMs of coagulum detected in Case 9 (surfactant-rich) are on average a  factor $1/2$ lower than the ones in Case 8 ($10-20$\% lower in surfactant coverage).
On the other hand, the right hand side of the same figure shows a steep increase of detected coagulum between the end of the feed additions and the end of the polymerization in Case 9. Indeed, we have measured a value almost three times higher with respect to the value in Case 8 at the same time step in the process.\\
\begin{figure}
	\centering
	\subfloat{\includegraphics[width = 0.47 \linewidth]{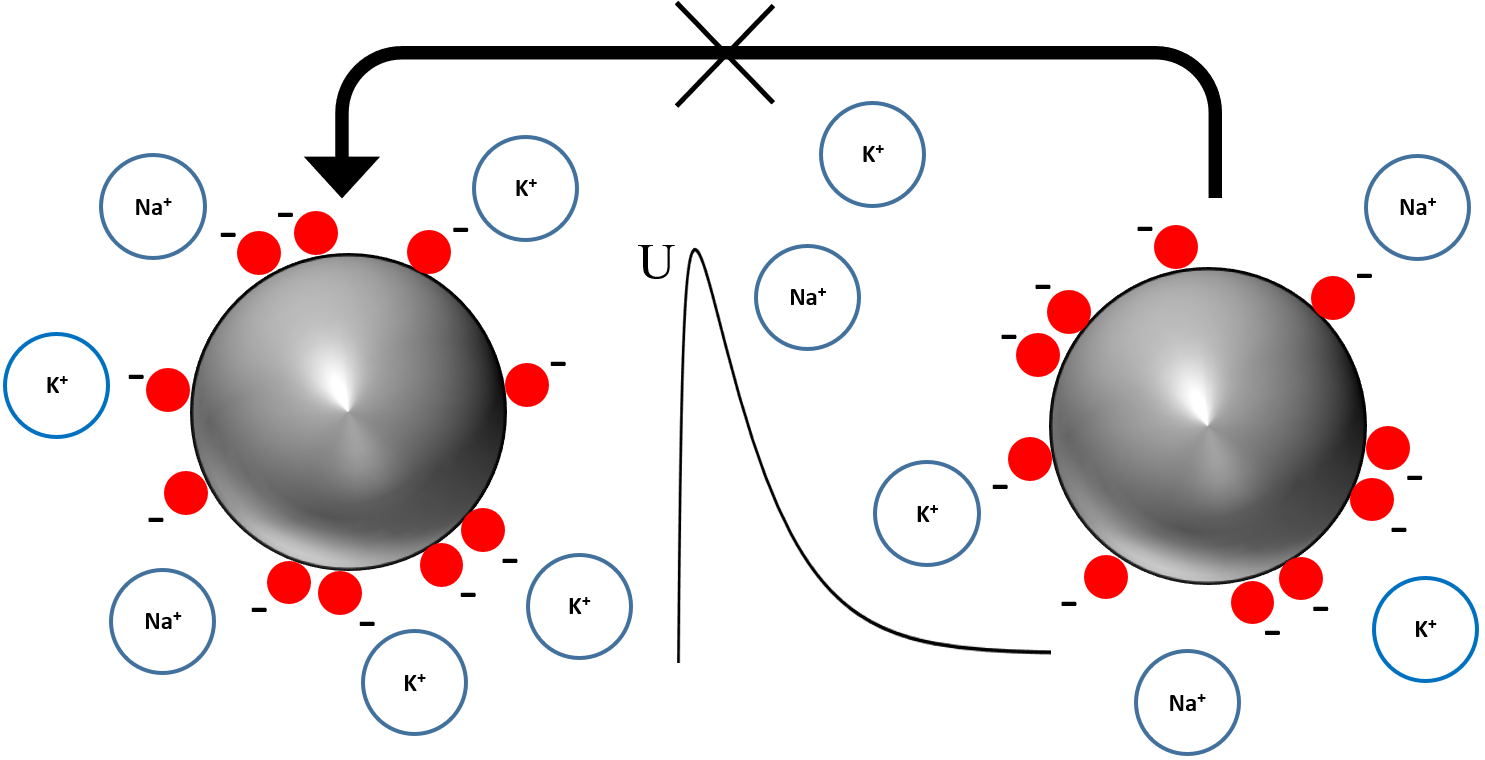}} \quad
	\subfloat{\includegraphics[width = 0.46 \linewidth]{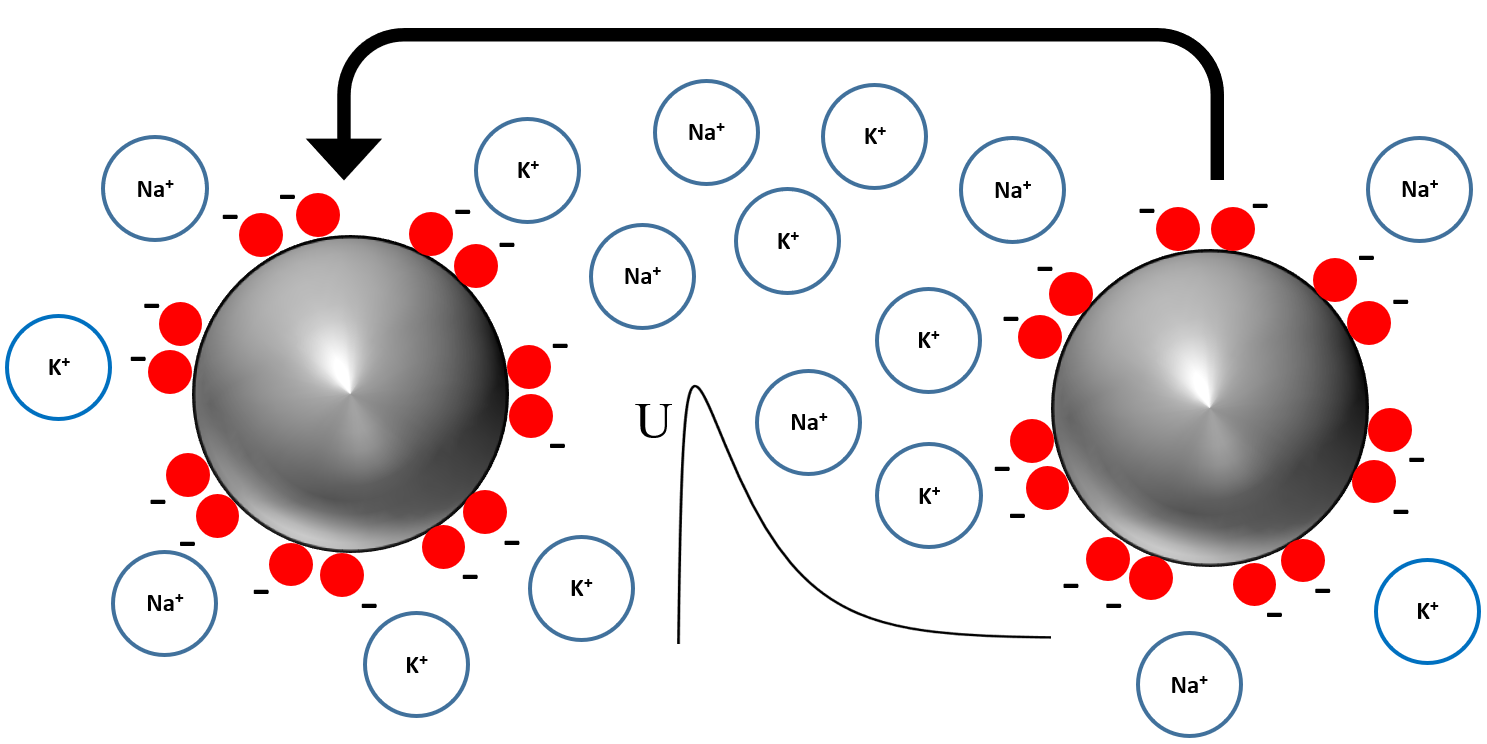}}
	\caption{Visual summary of the balance between the surface coverage and salt content on the energy barrier against aggregation in Case 8 (left) and Case 9 (right) at the end of the polymerization.}
	\label{CCC_Cartoon}
\end{figure}
The explanation behind this behavior has been visually summarized in Figure~\ref{CCC_Cartoon}: it is very likely that the concentration of counterions in Case 9 crosses the Critical Coagulation Concentration (CCC) of the system between the end of the feed additions and the end of the process. So, even if Case 8 has a lower surface coverage the salt content is low enough to guarantee a sufficiently high energy barrier which limits the aggregation of the particles.
On the other hand, since Case 9 probably passes the CCC the energy barrier will be overall lower than in Case 8, even if the surface coverage is higher, because of the stronger screening effect on the electrical double layer caused by the counterions. This ultimately  triggers the colloidal instability and sudden formation of coagulum; it can be observed how this balance is very delicate since an increase of salt concentration of 7\% leads to an increase of 300\% in presence of coagulum.\\
From inspection of Figure \ref{Cationic_Concentration}, it is hypothesized that the critical point is associated for a value of [CI] between $ 0.085 \ \text{and} \ 0.09$ mol/L. This value is lower than the ones reported in the literature for carboxylic latexes in alkaline environments: with $\text{Na}^+$ or $\text{K}^+$ as counterions the reported CCC is $\approx$ 0.35 mol/L \cite{Ehrl2009}, which is significantly higher than the predicted one. \\
A possible explanation for this discrepancy relies on the fact that the dominant counterion in our system is the ammonium ion $\text{NH}_{4}^{+}$, which have a more destabilizing effect on the dispersion, compared to $\text{Na}^+$ or $\text{K}^+$ according to the Hofmeister series \cite{Marte}. Indeed, within the Hofmeister, a difference of a factor $2-3$ upon changing the counterion is not uncommon. 
Furthermore, the CCCs reported across the literature are typically related to extremely dilute conditions, whereas in industrial conditions the colloidal particle concentration is much higher such that many-body effects can enhance coagulation on top of EDL screening effects.
It is evident that this system is very sensitive to increase in salt content even far away from the typical CCC values because of (i) the stronger destabilizing effect of the ammonium ion, and (ii) the much higher solids content typical of the industrial emulsion polymerization processes.

\section{Conclusions \& future steps}
In this work we developed a mathematical model to predict the seeded-emulsion polymer reaction kinetics of co- and ter-polymerizations. The model combines the Smith-Ewart equations with state-of-art models for radical exchange between particle and aqueous phase within the pseudo-homopolymerization framework. Unknown parameters related to the monomer-specific influence on the gel effect are calibrated by fitting literature data of monomer conversion for homo- and co-polymerizations. This leads to a predictive model for the seeded-emulsion ter-polymerization of  \textit{n}-butyl acrylate (n-BA) and methyl methacrylate (MMA) with 2-hydroxyethyl methacrylate (2-HEMA), with sodium persuphate as the initiator. The model predictions are compared with DLS and GC characterizations of the time-evolution of average particle size and the conversion, and good agreement is found with no adjustable parameters. This quantitative model for the particle size evolution is then combined with the two-step surfactant adsorption isotherm appropriate for latex particles and ionic surfactants, and with the relevant chemical association equilibria for the various species present in solution (including buffers, etc) to predict the particle surfactant surface coverage and the total concentration of counterions throughout the entire polymerization process. The methodology is applied to two industrial test cases of n-BA/MMA/2-HEMA ter-polymerization that were carried out with different amounts of the same ionic surfactant. The model analysis shows that the surfactant-rich system displays significantly less coagulation (better colloidal stability) during all steps of the industrial polymerization process except for the last step, which implies additions of ammonia to control the pH. These ammonia additions clearly drive the \textit{total} counterion concentrations to higher values compared to the system with less surfactant. This is likely to bring the system above the critical coagulation concentration (CCC) and to uncontrolled coagulation resulting in a much larger amount (by a factor three) of detected coagulum at the end of the process. 
Hence, the proposed modelling-based methodology offers the possibility to quantitatively rationalize the interplay of surfactant and counterion concentrations on colloidal coagulation during emulsion poylmerization. This, in turn, opens up the way to achieving optimal control over coagulation in industrial and lab-scale emulsion polymerization processes.
In future work, the use of the model to quantify the surfactant surface coverage and the total ionic strength of the system will serve as a starting point for a systematic quantitative evaluation of the Fuchs stability ratio through Eq.(\ref{Fuchs_Stability_Ratio}), including also possible loss of surface charge due to association between counterions and surface charge groups~\cite{Zichen}, and the effect of shear flow and hydrodynamic interactions~\cite{Zaccone2009}.
\begin{acknowledgement}
L.B. gratefully acknowledges financial support from Synthomer UK Ltd. We would like to thank Massimo Morbidelli for many useful discussions and input.
\end{acknowledgement}

\begin{suppinfo}
\subsection{Extensive description of the monomer partitioning scheme}
The monomer partitioning is a procedure whose ultimate goal is the evaluation of the following properties:
\begin{enumerate}
	\item Volume fraction of each monomer in particle $\phi_j^p$, aqueous $\phi_j^w$ and droplet phase $\phi_j^d$;
	\item Volume fraction of polymer in the particle phase $\phi_\text{pol}^p$;
	\item Volume fraction of water in the aqueous phase $\phi_\text{water}^w$;
	\item Total volumes of particle $V^p$, aqueous $V^w$ and droplet $V^d$ phase.
\end{enumerate}
through the knowledge of
\begin{enumerate}
	\item the volume of water \textit{W} from the formulation;
	\item the volume of the growing polymer phase $V_\text{pol} = m_\text{P}/\rho_\text{P}$ known by the knowledge of the initial mass of the seed and the total mass of the added polymer known by Eq.(\ref{Polymer_Mass_Balance});
	\item the partition coefficients $K_j^k$.
\end{enumerate}
 We need to determine 3 $N_m$ + 5 variables, so we need the same number of equations which are formally defined as
\begin{enumerate}
	\item Conservation of the volume fraction of each j-th monomer $\phi_j^k$ in every phase (+3);
	\item Volume balance of water (+1);
	\item Volume balance of polymer (+1);
		\item Volume balance of each monomer ($N_m$);
	\item Partitioning of each monomer between droplet-aqueous phase and particle-aqueous phase (2 $N_m$).
\end{enumerate}
The full algebraic system to be solved at each time can be written as
\begin{equation}
\begin{cases}
\sum_{j=1}^{\text{N}_m} \phi_j^p + \phi_\text{pol}^p  = 1;  \\
\sum_{j=1}^{\text{N}_m} \phi_j^d = 1; \\
\sum_{j=1}^{\text{N}_m} \phi_j^w + \phi_\text{water}^w = 1; \\
V_j = \phi_j^p V^p + \phi_j^d V^d + \phi_j^w V^w, \ j =1,...,N_m; \\
\phi_\text{water}^w V^w = W; \\
\phi_\text{pol}^p V^p = V_\text{pol}; \\ 
K_j^d = \dfrac{\phi_j^d}{\phi_j^w}, \ j =1,...,N_m; \\
K_j^p = \dfrac{\phi_j^w}{\phi_j^w}, \  j =1,...,N_m. \\ 
\end{cases} 
\label{Monomer_Partitioning}
\end{equation}
\subsection{State dependent radical desorption rates $\textbf{k}_\textbf{i}^\textbf{m}$}
The state dependent desorption rate of each monomer $R_\text{dm,ij}'$ can be written as a function of the rate of appearance of monomeric radicals and the probability $Q_i^j$ for them to subsequently desorb:
\begin{equation}
	R_\text{dm,ij}' = \biggl[ \biggl(\sum_{k \in N_r'}  k_{\text{fm},kj} P_k [\text{J}]_\text{p} \biggr) i [N_i]  + \rho_{\text{re},i} [N_{i-1}] \biggr] Q_i^j,
\end{equation}
where [$N_i$] is the number concentration of particles with state i which represents the total number of propagating chains inside it.\\
The first term represents the contribution from the chain tranfer, while the second the re-entry of a radical previously desorbed.
We will write the re-entry rate $\rho_\text{re,k}$ as a function of a state-average desorption coefficient $\langle k^j \rangle$, the average number of radicals inside each particle $\tilde{n}$ and the aforementioned fate parameter $\beta_j$:
\begin{equation}
	\rho_\text{re,j} = \langle k^j \rangle \tilde{n}(1-\beta_j).
\end{equation}
At this point we can write the the desorption frequency rate as
\begin{equation}
	R'_\text{dm,ij} = k_i^j i [N_i] = \biggl( \sum_{k \in N_r'} k_{\text{fm},kj}  P_k [\text{J}]_\text{p} i [N_i]  + \rho_{\text{re},j} [N_{i-1}] \biggr) Q_i^j
\end{equation}
which means
\begin{equation}
	k_i^j = \biggl[ \sum_{k \in N_r'}   k_{\text{fm},kj} P_k [\text{J}]_\text{p} + \biggl( \dfrac{\langle k^j \rangle \tilde{n}(1-\beta_j) [N_{i-1}]}{ i [N_i]} \biggr)\biggr] Q_i^j
\end{equation}
In order to find $\langle k^j \rangle$ we need to explicitly write the overall rate of desorption for each monomer as
\begin{equation}
	R'_\text{dm,tot,j} = \langle k^j \rangle \tilde{n} [N_T] = \sum_{i} R'_\text{dm,ij} = \sum_{i} \biggl( \sum_{k \in N_r'}  k_{\text{fm},kj} P_k [\text{J}]_\text{p} i [N_i]  + \langle k^j \rangle \tilde{n}(1-\beta_j) [N_{i-1}] \biggr) Q_i^j
\end{equation}
where $[N_T]$ is the total concentration of particles inside the reactor; from the previous equation it is then possible to find the state average desorption coefficient as
\begin{equation}
	\langle k^j \rangle = \biggl( \sum_{k \in N_r'} k_{\text{fm},kj} P_k [\text{J}]_\text{p} \biggl) \dfrac{\sum_{i} i N_i Q_i^j }{\tilde{n}[ 1 - (1-\beta_j) \sum_{i}N_{i-1}Q_i^j]}
\end{equation}
by also introducing the probability $N_i$ to find a particle in state \textit{i} as
\begin{equation}
	N_i = \dfrac{[N_i]}{[N_T]}.
\end{equation}
\subsection{Approximate analytic solution to the Maxwell-Morrison mechanism for co- and ter-polymerizations}
The begin of the "control by aqueous phase growth" mechanism is the decomposition of initiator which leads to the formation of radical precursors which start reacting with the i-th monomeric species. This process triggers the formation of oligomeric radicals which can (i) keep propagating with other monomers or (ii) terminate with another chain whose total concentration is [\ce{T^.}]. Once they reach a critical degree of oligomerization, namely \textit{z}, they have become sufficiently surface active to instantaneously migrate to a particle and enter it with rate $\rho_I$.\\
The most crucial step is the description of the propagation: according to the Pseudo-Homopolymerization approach the total propagation rate in a certain phase \textit{o} is calculated by adopting the same formalism as in a homopolymerization, but through average propagation rate constants related to the consumption for each monomer \textit{j} $\overline{k}_{\text{p},j}$ calculated as a weight average among every rate constant involving \textit{j} and any reactive site \textit{i} it can interact with and the probability $P_i$ that each chain owns that particular terminal end:
\begin{equation}
R_\text{p,o} = \sum_{i=1}^{N_m} \biggl( \sum_{j=1}^{N_r} \overline{k}_{\text{p},ji} P_j \biggr)  [i]_o = \sum_{i=1}^{N_m}\overline{k}_{\text{p},i} [i]_o 
\end{equation}
Unfortunately, since the long chain approximation is not respected in the aqueous phase, this approach can not be used to compute the average propagation rates $\overline{k}_{\text{p},j}^\text{w}$ in the aqueous phase. For this reason we have decided to evaluate them by adopting a geometric average among all the possible propagation rates involving that particular monomer and any possible active site: 
\begin{equation}
k_{\text{p},i}^\text{w} = \biggl( \prod_{j=1}^{N_r} k_{\text{p},ji} \biggr)^{1/N_r}
\label{Average_Propagation_Rates_Aqueous}
\end{equation}
Now it is possible to write the reacting scheme for the propagating chains in the aqueous phase with degree of polymerization \textit{k} independently from the type of active site \textit{k}: $\ce{R^._{\text{k,tot}}} = \sum_{j=1}^{N_r} \ce{R^._{\text{k,j}}}$. 
\begin{equation}
\begin{cases}
\ce{I ->[f k_{\text{d}}] 2 \ce{I^.}} \\
\ce{I^. + i ->[k_{\text{p},Ii}] R^._{1,i}} \quad \text{i} = 1, ..., N_m \\
\ce{R^{.}_{k,tot}  + i ->[\overline{k}_{\text{p},i}] \ce{R}^{.}_{k+1,tot}} \quad 1 \leq \text{k} < z \ \ \text{i} = 1, ..., N_m \\
\ce{R^{.}_{k,tot}  + T^{.} ->[\langle k_{\text{t,w}}\rangle] Dead Chains} \quad 1 \leq \text{k} < z \\
\ce{R^{.}_{z,tot}  + Particle ->[\rho_{I}] Entry}
\end{cases}
\label{Entry_Reacting_Scheme}
\end{equation}
From Eq.(\ref{Entry_Reacting_Scheme}) it is possible to write down the balance for every \ce{R^._{\text{k,tot}}}
\begin{equation}
\begin{cases}
\dfrac{d \ce{I^.}}{dt} = 2 f k_\text{d} \ce{I} - \sum_{i=1}^{N_m} k_{\text{p},Ii}[\text{i}]_\text{w} \ce{I^.} \\ \\
\dfrac{d \ce{R^.}_{1,\text{tot}}}{d t} = \sum_{i=1}^{N_m} k_{\text{p},Ii} [\text{i}]_\text{w} \ce{I^.} - \sum_{i=1}^{N_m} \overline{k}_{\text{p},i}^\text{w} [\text{i}]_\text{w} \ce{R^.}_{1,\text{tot}} - 2 \langle k^\text{t}_\text{w} \rangle [\ce{T^.}] \ce{R^.}_{1,\text{tot}} \\ \\
\dfrac{d \ce{R^.}_{k,\text{tot}}}{d t} = \sum_{i=1}^{N_m} \overline{k}_{\text{p},i}^\text{w}  [\text{i}]_\text{w} \ce{R^.}_{\text{k}-1,\text{tot}} - \sum_{i=1}^{N_m} \overline{k}_{\text{p},i}^\text{w} [\text{i}]_\text{w} \ce{R^.}_{\text{k,tot}} - 2 \langle k^\text{t}_\text{w} \rangle [\ce{T^.}] \ce{R^.}_{\text{k,tot}} \\ \\
\dfrac{d \ce{R^.}_{z,\text{tot}}}{d t} = \sum_{i=1}^{N_m} \overline{k}_{\text{p},i}^\text{w}  [\text{i}]_\text{w} \ce{R^.}_{z-1,\text{tot}} - \rho_I \dfrac{N_\text{P}}{N_\text{AV}}	
\end{cases}
\end{equation}
Under the steady state approximation it is possible to find the following expressions for the different \ce{R^{.}_{\text{k,tot}}}:
\begin{equation}
\begin{cases}
\ce{R^._{1,\text{tot}}} = \dfrac{2 f k_\text{d} \ce{I}}{\sum_{i=1}^{N_m} k_{\text{p},Ii}[\text{i}]_\text{w} \ce{I^.} + 2 \langle k^\text{t}_\text{w} \rangle [\ce{T^.}] } \\ \\
\ce{R^._{k,tot}} = \dfrac{\sum_{i=1}^{N_m} \overline{k}_{\text{p},i}^\text{w}[\text{i}]_\text{w}}{\sum_{i=1}^{N_m} \overline{k}_{\text{p},i}^\text{w} [\text{i}]_\text{w} + 2 \langle k^\text{t}_\text{w} \rangle [\ce{T^.}]} \ce{R^._{k-1,tot}} =  \biggl( 1 + \dfrac{2 \langle k^\text{t}_\text{w} \rangle [\ce{T^.}]}{\sum_{i=1}^{N_m} \overline{k}_{\text{p},i}^\text{w} [\text{i}]_\text{w}}\biggr)^{-1} \ce{R^._{k-1,tot}}
\end{cases}	
\end{equation}
Finally, if we suppose the mixture of monomers to be sufficiently hydrophobic, then the total concentration of radical chains [\ce{T^.}] can be approximated as
\begin{equation}
[\ce{T^.}] \sim \sqrt{\dfrac{ f k_\text{d}[\ce{I}]_\text{w}}{\langle k_\text{w}^\text{t}\rangle } }
\end{equation}
and it is possible to obtain the following modified version of the original control by aqueous phase growth which is reported in Eq.(\ref{Entry_Rate_Maxwell_Morrison}):
\begin{equation}
\rho_\text{I} = \dfrac{2 f k_\text{d} \text{I} N_\text{AV}}{N_\text{p}} \biggl( \dfrac{2 \sqrt{f k_\text{d} [\text{I}]_\text{w} \langle k_\text{w}^\text{t} \rangle}}{\sum_{i=1}^{N_m}\overline{k}_\text{p,i}^\text{w} [\text{i}]_\text{w}}  + 1 \biggr)^{1-\overline{z}}
\label{Entry_Rate_Maxwell_Morrison_SI}
\end{equation}
Note that Eq.(\ref{Entry_Rate_Maxwell_Morrison}) has been modified by adopting a variable critical length critical length $\overline{z}$ already introduced in the main text which will be defined as
\begin{equation}
	\overline{z} = \sum_{i=1}^{N_m} \dfrac{\phi_i^w}{\sum_{j=1}^{N_m} \phi_j^w} z_i,
\end{equation}
to take into account the different relative presence of the various monomers which plays an impact on when the produced oligomers will become surface active enough to interact with the particle phase.\\
Every $z_i$ has been chosen as an intermediate value between the minimum degree of polymerization for surface activity and the one which will cause incipient water insolubility according to the following expressions derived from thermodynamic considerations \cite{Maxwell} :
\begin{equation}
1 + \text{int}\biggl( \dfrac{ -23 [\text{kJ/(mol K)}]}{RT \ln{[\text{i}]_\text{w,sat}}}\biggr) \le z_i \le 1 + \text{int}\biggl( \dfrac{ -55 [\text{kJ/(mol K)}]}{RT \ln{[\text{i}]_\text{w,sat}}}\biggr),
\label{Homopolymerization_DOPs}
\end{equation}
where $[\text{i}]_\text{w,sat}$ is the molar concentration of every monomer at saturation conditions and int is a function which approximates the value in the brackets to the smallest integer; the final values adopted in this work are reported in Table \ref{Extra_Inpout_Data}.
\subsection{Dynamic Light Scattering procedure}
The average particle sizes have been evaluated through \textit{Dynamic Light Scattering} (DLS) using a Malvern ZetaSizer.\\
Since the solid content in the samples is really high (up to 50\%) the measurements can not be run  after sampling, but a procedure of successive dilutions is requested.
In this way it is possible to obtain a good quality for the signal from the DLS in order to achieve reliable values for the average particle size which have been proposed in Fig.\ref{BN21303_kinetic_variables}.\\
With this proceudre it is also possible to confirm the hypothesis of monodispersity of the population introduced at the beginning of the paper by proposing in Table \ref{Table:Polydispersity_Index} the temporal trend of the Polydispersity Index (PDI) for Case 9: since the values have an order of magnitude of $10^{-2}$ it can be confirmed that all of the particles have basically the same particle size; for Case 8 it is available only the PDI for the final sample which has been measured to be 0.011, so in line with the values proposed in Table~\ref{Table:Polydispersity_Index}.
\begin{table}
	\centering
	\begin{tabular}{|c|c|c|}
		\hline 
		\textbf{Time [min]} & 	\textbf{Case 9}  \\
		30          	    & 		0.02	   \\
		120         	    & 		0.002	   \\
		150          	    & 		0.016	   \\
		210             	& 		0.015	   \\
		345                 &  		0.053	   \\
		\hline
	\end{tabular}
	\caption{Temporal trend of Polydispersity Indexes for IC 1 and Case 9 with time = 0 set as the beginning of the monomers' additions.}
	\label{Table:Polydispersity_Index}
\end{table}
\subsection{Impact of the backbiting on the kinetic variables of BA/MMA copolymerizations}
\begin{figure}
	\centering
	\subfloat{\includegraphics[width = 0.7 \linewidth]{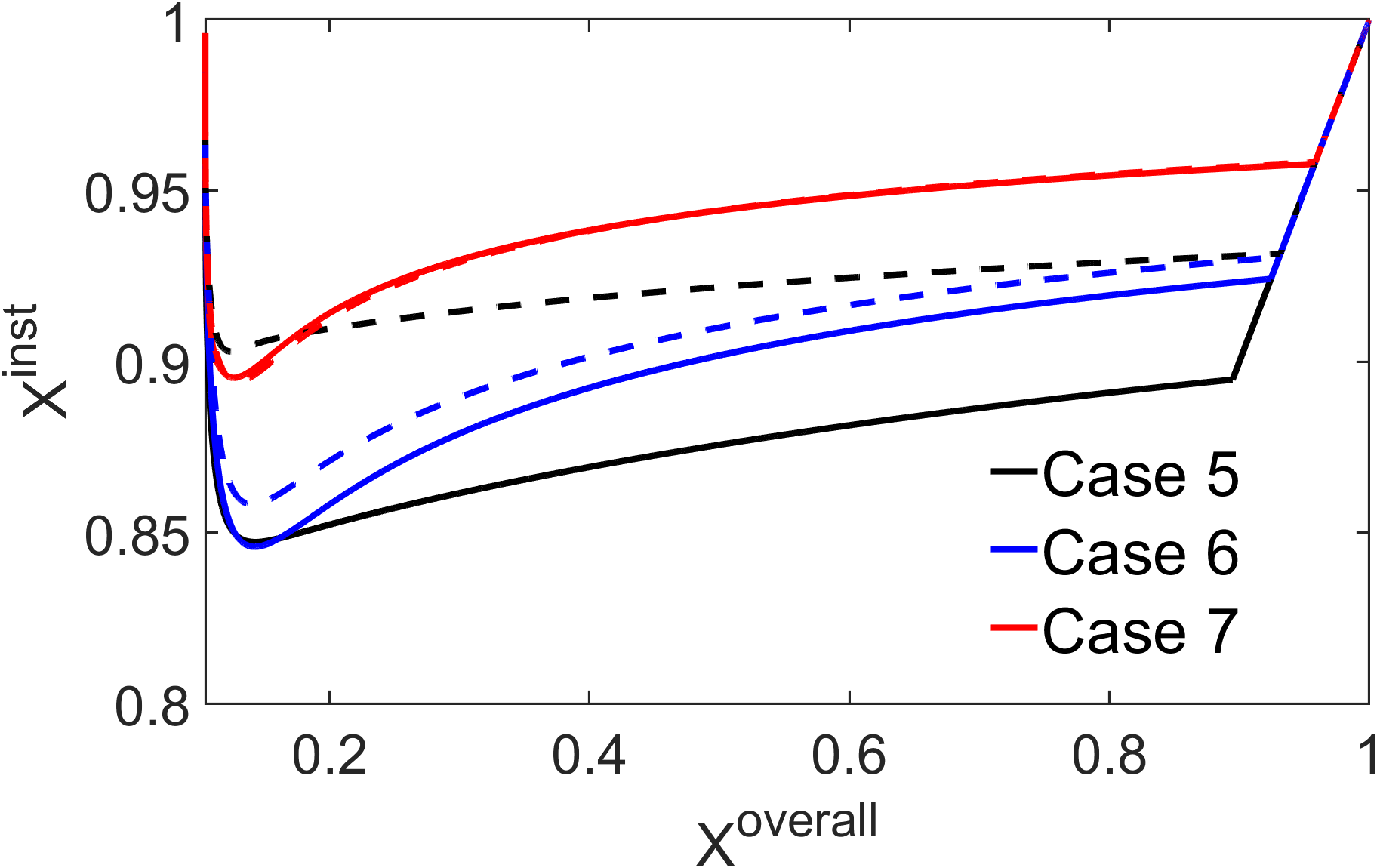}} \quad
	\caption{Plots of Instantaneous vs Overall conversion for the co-polymerization test cases considering (solid lines) and neglecting (dashed lines) the presence of tertiary radicals in the reacting scheme.}
	\label{Backbiting_Effect}
\end{figure}
This small section has then the intent of analysing the effect provided by the backbiting on the three test cases considered for the co-polymerizations (Cases 5, 6 and 7).\\
The absence of MCRs in the particle phase has been simulated by imposing $k_\text{fp,2} = 0$, meanwhile in the aqueous phase their presence has been neglected by considering $k_\text{t,w}^\text{BA} = k_\text{t,11}$ (it is not an adjustable parameter anymore) and not including $k_\text{p,21}$, the propagation rate constant for the MCRs in the evaluation of the average propagation rate in the particle phase computed by Eq.(\ref{Average_Propagation_Rates_Aqueous}).\\
In Fig.\ref{Backbiting_Effect} it can be seen that there is an important difference between neglecting and considering the presence of MCRs for Case 5 only in which the mole fraction of methyl methacrylate is 0.1, meanwhile there is a negligible difference on the trends related to Case 6 and Case 7, where the mole fraction of MMA is 0.3 and 0.5 respectively.
By analysing the pieces of literature which focus on the role of the backbiting in co-polymerizations \cite{hlalele1,hlalele2} it emerges that the experimental data, which have been gathered from a \textit{n}-BA/styrene copolymer, have been obtained by considering a minimal mole fraction of styrene of 0.3, the lower limit for which the model provides with almost identical results considering or neglecting the presence of tertiary radicals.\\
Concluding, the data reveal that the backbiting and the mid-chain radicals still play an important role in the polymerization kinetics of a copolymer involving really high mole fractions of n-Butyl Acrylate, meanwhile its effect can be neglected in case of mole fraction of the second monomer higher than 0.3 as it has been demonstrated experimentally.
\end{suppinfo}

\bibliography{achemso}

\end{document}